\documentclass[sigconf]{acmart}

\usepackage{amssymb}

\AtBeginDocument{%
  }

\setcopyright{acmlicensed}
\acmYear{2026}\copyrightyear{2026}
\acmConference[ASIA CCS '26]{ACM Asia Conference on Computer and Communications Security}{June 1--5, 2026}{Bangalore, India}
\acmBooktitle{ACM Asia Conference on Computer and Communications Security (ASIA CCS '26), June 1--5, 2026, Bangalore, India}
\acmDOI{10.1145/3779208.3807832}
\acmISBN{979-8-4007-2356-8/26/06}

\usepackage{tikz}
\usetikzlibrary{shapes.geometric, arrows, positioning, fit, calc, shapes}
\usepackage{amsmath}
\usepackage{hyperref}
\usepackage[table]{xcolor}
\usepackage{booktabs}
\usepackage{graphicx}
\usepackage{amssymb}
\usepackage{pgfplots}
\pgfplotsset{compat=1.18}
\usepackage{adjustbox}
\usepackage{enumitem}
\usepackage{soul}
\usepackage{multicol}
\usepackage{diagbox}
\usepackage{cleveref}
\usepackage[most]{tcolorbox}
\usepackage{array}

\microtypecontext{spacing=nonfrench}

\usepackage{algorithm}
\usepackage{algpseudocode}
\algrenewcommand\algorithmiccomment[1]{\hfill{\scriptsize\textit{// #1}}}

\newtcolorbox{resultbox}[1][]{
  colback=gray!5,colframe=gray!40!black,breakable,title=#1
}

\definecolor{minor}{RGB}{204,255,204}
\definecolor{moderate}{RGB}{255,242,204}
\definecolor{high}{RGB}{255,179,186}

\newcommand{\qrate}{q}
\newcommand{\perturb}{\ensuremath{\qrate}}            
\newcommand{\dpnoise}{\ensuremath{\varepsilon_{\mathrm{DP}}}}

\newcommand{\riskTotal}{29}
\newcommand{\riskIdentCount}{7}

\newcommand{\riskIdent}{0.241} % 7/29
\newcommand{\riskMeta}{0.138}  % 4/29
\newcommand{\riskBias}{0.414}  % 12/29
\newcommand{\riskNet}{0.345}   % 10/29

\newcommand{\riskNorm}{0.606}

\begin{document}

\title{SoK: Analysis of Privacy Risks and Mitigation in Online Propaganda Detection through the PROMPT Framework}

\author{Dhiman Goswami}
\affiliation{%
 \institution{George Mason University}
 \city{Fairfax}
 \state{Virginia}
 \country{United States of America}}
\email{dgoswam@gmu.edu}

\author{Al Nahian Bin Emran}
\affiliation{%
 \institution{George Mason University}
 \city{Fairfax}
 \state{Virginia}
 \country{United States of America}}
\email{abinemra@gmu.edu}

\author{Md Hasan Ullah Sadi}
\affiliation{%
 \institution{George Mason University}
 \city{Fairfax}
 \state{Virginia}
 \country{United States of America}}
\email{msadi@gmu.edu}

\author{Sanchari Das}
\affiliation{%
 \institution{George Mason University}
 \city{Fairfax}
 \state{Virginia}
 \country{United States of America}}
\email{sdas35@gmu.edu}
%\author{Anonymous}
\renewcommand{\shortauthors}{Goswami et al.}

\begin{abstract}
Online propaganda detection pipelines expose measurable privacy risks at multiple stages including data collection, feature extraction, and model inference. We conduct a structured analysis of $162$ peer-reviewed studies and formalize the problem using the \textit{Propaganda Risk Online Mitigation and Privacy-preserving Tactics} (PROMPT) framework. PROMPT models risks $R$ and mitigation strategies $S$ through a mapping $\mathcal{M}:R\to S$ guided by a utility function $\alpha\cdot \mathrm{PrivacyGain}(s_j) - \beta\cdot \mathrm{PerfLoss}(s_j) - \gamma\cdot \mathrm{Cost}(s_j)$, with tunable $(\alpha,\beta,\gamma)$ enabling stakeholders to balance privacy, accuracy, and deployment costs. To assess practical adoption, we introduce a compliance score that quantifies the alignment of existing methods with GDPR, CCPA etc. requirements. Our evaluation shows that many widely used pipelines remain non-compliant, particularly in metadata handling and user-level aggregation. We further present empirical fine-tuning experiments on transformer-based encoders and decoders under synthetic perturbation, demonstrating a monotonic privacy–utility trade-off: with $q = 0.05$ performance decreased by 1--2\% F$_1$, while at $\perturb=0.20$ the reduction reached 13--14\%. These results establish quantitative baselines for privacy costs in propaganda detection. Our contributions include a formal risk-to-defense mapping, a compliance-oriented auditing metric, and experimental evidence of privacy–performance trade-offs, providing a technical foundation for building regulation-compliant and privacy-aware detection systems.
\end{abstract}





\maketitle

\section{Introduction}  
In the digital age, individuals increasingly rely on platforms such as Meta (formerly Facebook), X (formerly Twitter), YouTube, and major news outlets like BBC, CNN, and Sky News to share and consume information~\cite{newman2024reuters}. While these platforms enable global connectivity and real-time access to news, they also act as fertile grounds for the rapid spread of \textit{propaganda.} Manzoor et al. define propaganda as \textit{\lq\lq a deliberate, conscious, malicious, and cunning effort by a group, organization, or individual to control and influence public beliefs and actions through selected truth, mass communication, and personal contact\rq\rq}~\cite{manzoor2019propaganda}. Prior studies further show that propaganda spreads faster and more widely than factual information on social media~\cite{patel2017modeling,vosoughi2018spread,surani2026co,sharevski2024debunk,sharevski2025social}.

The vast scale of user-generated content and its rapid circulation make effective detection increasingly challenging~\cite{horak2021technological,kumar2025privacy}. Propaganda transcends linguistic and platform boundaries~\cite{ng2021does, saleh2023beneath}, while human cognitive biases and engagement-driven algorithms exacerbate its reach~\cite{meserole2018misinformation, maseri2020socio}. To address these challenges, researchers employ machine learning (ML) models such as BERT, RoBERTa, and GPT-based architectures to detect common techniques including framing, persuasion, and disinformation~\cite{krak2024method, piskorski2023semeval}.  

Progress has been enabled by curated datasets capturing political, military, and commercial disinformation~\cite{harris2024fake}, such as AG-News, ArAIEval, the DARPA Twitter Bot Challenge, and Kaggle's fake news dataset. Multilingual corpora like EUvsDisinfo highlight state-sponsored campaigns, e.g., pro-Kremlin narratives~\cite{leite2024euvsdisinfo}. Yet, despite their utility, such resources raise serious privacy concerns~\cite{el2024preserving}, as large-scale data collection introduces risks of surveillance, re-identification, and misuse~\cite{santos2023artificial, saheb2023ethically}. Detection techniques that exploit behavioral or biometric signals may further infringe on individual rights~\cite{saheb2023ethically}. While privacy-preserving approaches such as DP, anonymization, and secure multi-party computation have been explored~\cite{semantha2023pbdinehr,li2020review}, most studies prioritize accuracy, leaving privacy and compliance under-addressed~\cite{martino2020survey,semantha2023pbdinehr,saka2025sok,adhikari2023evolution,khadka2026sok,tazi2025multi,podapati2025sok}.

To examine this gap, we analyze $162$ peer-reviewed studies spanning datasets, ML methodologies, and privacy-preserving technologies. Building on this analysis, we propose the \textit{Propaganda Risk Online Mitigation and Privacy-preserving Tactics} (PROMPT) framework, which maps privacy threats (e.g., metadata leakage, behavioral inference) to mitigation strategies (e.g., DP, federated learning, secure multi-party computation) and integrates regulatory principles and best practices to guide the design of performant, privacy-preserving, and compliant detection systems. Our study is structured around the following research questions:

\begin{itemize}[leftmargin=*]
    \item~\textbf{RQ1:}~\textit{What privacy risks arise during data collection, model training, and deployment in propaganda detection?} \label{rq:1}

    \item~\textbf{RQ2:}~\textit{How well do current propaganda detection methods incorporate privacy-preserving techniques, and how effectively do they mitigate risks across different data modalities, platforms, and languages?} \label{rq:2}

    \item~\textbf{RQ3:}~\textit{What privacy-preserving methods can be integrated into propaganda detection systems to protect user data while maintaining detection performance?} \label{rq:3}
\end{itemize}

\noindent Our work offers the following key~\textbf{contributions}:

\begin{itemize}[leftmargin=*]

\item Here we report on the first SoK on the intersection of propaganda detection and privacy, based on a systematic analysis of $N=162$ peer-reviewed studies covering datasets, detection methods, and privacy-preserving techniques.
Let $F=\{d,m,p\}$ denote features for datasets $(d)$, methods $(m)$, and privacy techniques $(p)$.  
The adoption ratio is defined as
\[
C_x \triangleq \frac{|P_x|}{N}, \quad x\in\{d,m,p\},
\]
providing a reproducible baseline to compare uptake across areas (e.g., metadata use, privacy-technique adoption).  

\item We develop the \textit{PROMPT} framework as a mapping $\mathcal{M}: R \rightarrow S$, from risk categories $R=\{r_1,\dots,r_k\}$ (e.g., re-identification, metadata leakage) to mitigation strategies $S=\{s_1,\dots,s_\ell\}$ (e.g., DP, SMPC, FL).  
Defense selection is posed as utility optimization:
\[
s^*(r_i) \in \arg\max_{s_j \in S}
\ \Big[\ \alpha\,\mathrm{PrivacyGain}(s_j) - \beta\,\mathrm{PerfLoss}(s_j)\ \Big],
\]
with tunable $(\alpha,\beta)$ reflecting deployment priorities.  

\item For $M$ evaluated pipelines, we define the compliance score
\[
\mathrm{CompScore} \triangleq \frac{1}{M}\sum_{i=1}^{M} \mathbf{1}\!\left[\textsc{Compliant}(\mathrm{Method}_i)\right],
\]
and compile a best-practice dictionary $\mathcal{B}$ linking regulatory clauses (e.g., GDPR data minimization, PbD) to technical controls (e.g., DP noise calibration, $k$-anonymity, metadata filtering), offering actionable guidance.  

\item We empirically observe a monotonic privacy–utility trade-off with increasing perturbation rate $q$, i.e., $F_1(q)\le F_1(0)$ and $F_1(q_2)\le F_1(q_1)$ for $q_2 \ge q_1$ in our setup.

\end{itemize}

\section{Background}
In this section, we outline foundational concepts and techniques for propaganda detection on online platforms, emphasizing privacy risks when ML and Deep Learning (DL) models process user-generated content through data collection, model design, and NLP-based computation.

Propaganda has been categorized into techniques such as what-aboutism, straw man, red herring, bandwagon, reductio ad Hitlerum, exaggeration or minimization, thought-terminating cliché, causal oversimplification, appeal to fear or prejudice, black-and-white fallacy, name-calling or labeling, appeal to authority, loaded language, flag-waving, repetition, slogans, and doubt~\cite{da2019findings,stefan2024cyrillic,tomar2022araprop,chavan2022chavankane}. These categories underpin computational approaches, which are benchmarked at span, sentence, and document levels.

Early studies relied on traditional ML models such as Logistic Regression, SVM, Naïve Bayes, Decision Trees, and TF-IDF classifiers. Li et al.~\cite{li2019detection} applied a TF-IDF Logistic Regression with linguistic features to reduce identifiable data use, while Aggarwal and Sadana~\cite{aggarwal2019nsit} employed SVM with oversampling. Da San Martino et al.~\cite{da2019findings} integrated n-grams with LSTM and RoBERTa for multi-label detection, and infrastructure-level features were also leveraged to identify disinformation websites~\cite{hounsel2020identifying}.

Transformers advanced detection further.  
Kaczyński and Przybyła~\cite{kaczynski2021homados} proposed a BERT-based multi-task framework with credibility assessment, while Li et al.~\cite{li20211213li} fused RoBERTa and ResNet-50 for multimodal meme detection, raising privacy concerns in image-text fusion. Fadel et al.~\cite{fadel2019pretrained} showed BERT and USE ensembles outperform traditional ML, and Fouad and Weeds~\cite{fouad2024sussexai} applied AraBERT with augmentation for Arabic propaganda, emphasizing risks in under-resourced contexts. Broader surveys highlight methodological and ethical challenges in misinformation research~\cite{xiao2023sok}.

Other DL architectures also contributed. Blaschke et al.~\cite{blaschke2020cyberwalle} combined Bi-LSTM with BERT embeddings for token-level detection while minimizing metadata use. Gupta et al.~\cite{gupta2019neural} proposed a CNN-LSTM hybrid with segmentation and entity recognition to mitigate stylistic profiling. Gundapu and Mamidi~\cite{gundapu2022automatic} explored multimodal meme detection with CNN-LSTM, while Alhabashi et al.~\cite{alhabashi2024asos} extended this with ResNet-50 and MARBERT for Arabic memes, stressing re-identification risks. Patil et al.~\cite{patil2020bpgc} applied RNN-based sequence learning with BERT embeddings to news propaganda. User studies further reveal that misinformation warnings on video-sharing platforms may not always alter perceptions, complicating trust in detection systems~\cite{guo2023seeing}.

Recently, LLMs such as GPT-3.5, GPT-4, XLM-RoBERTa, and mT5 have been explored. Szwoch et al.~\cite{szwoch2024limitations} reported GPT-4’s inconsistencies and false positives, while Hamilton et al.~\cite{hamilton2024gpt} introduced GPT-assisted annotation for scalable, privacy-preserving labeling. Bagdasaryan et al.~\cite{bagdasaryan2022spinning} warned of \lq\lq propaganda-as-a-service\rq\rq via trigger phrases, and Morio et al.~\cite{morio2020hitachi} showed ensembles of XLNet, RoBERTa, and GPT-2 improve precision while reducing privacy exposure. Aldabbas et al.~\cite{aldabbas2025multiprop} introduced MultiProp, a cross-lingual LLM framework leveraging augmentation and meta-learning.

Despite these advances, significant privacy risks remain, including re-identification, metadata exposure, and profiling, particularly in cross-lingual and multimodal contexts involving large-scale user data and sensitive annotations.

\section{Method}
\begin{figure}[!ht]
    \centering
    \includegraphics[width=0.9\linewidth,height=5cm]{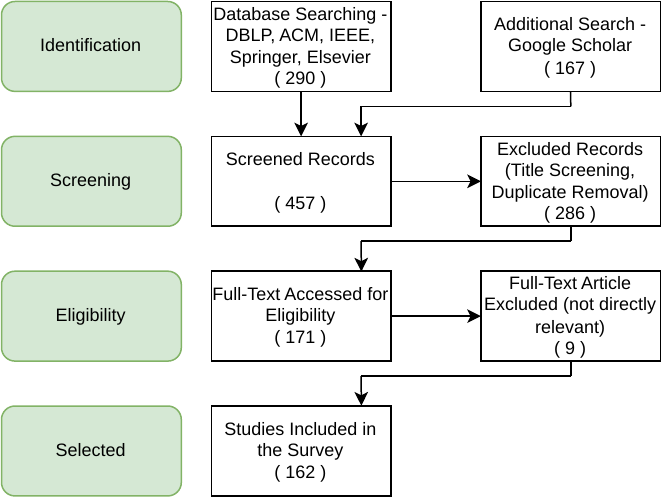}
    \caption{PRISMA Diagram: Stepwise Literature Collection, Filtering, and Selection Process.}
    \Description{PRISMA diagram Privacy Risk and Mitigation is Propaganda Detection SoK.}
    \label{fig:prisma}
\end{figure}

We follow the \textit{Preferred Reporting Items for Systematic Reviews and Meta-Analyses} (PRISMA) guidelines~\cite{mcinnes2018preferred} for conducting evidence-based literature review. Figure~\ref{fig:prisma} provides the overview of our study.

\subsection{Paper Identification}
We systematically searched high-impact academic databases, including \textit{DBLP, ACM Digital Library, IEEE Xplore, Springer, and Elsevier}, alongside top NLP, privacy, and HCI conferences (\textit{$A^{*}$ and $A$ venues} in CORE ranking~\footnote{https://portal.core.edu.au/conf-ranks/}). To ensure comprehensive coverage, we also retrieved relevant works from Google Scholar. 

\subsection{Paper Screening}
We designed structured queries to identify studies at the intersection of propaganda detection and privacy concerns. The primary keywords included: \textit{propaganda, online propaganda, propaganda privacy, propaganda mitigation}, with platform-specific terms such as \textit{propaganda in social media, propaganda in news media}. To capture privacy-preserving approaches, we included terms such as \textit{differential privacy, federated learning, anonymization in propaganda detection}. Despite rigorous filtering, potential biases in keyword and venue selection may have led to the exclusion of some studies. We included papers published until January $2025$. The final dataset provides a comprehensive analysis of privacy risks and mitigation strategies in propaganda detection.

\subsection{Paper Selection}
We retrieved $457$ research papers and applied a multi-stage filtering process aligned with best practices in systematic literature reviews. 

\textbf{Title and Abstract Screening:}  
First, we removed duplicates and irrelevant studies, retaining 286 unique paper. We then conducted manual abstract screening for relevance, excluding papers that lacked discussions on privacy risks, mitigation techniques, or ethical considerations. Three independent researchers conducted the screening and conducted a thematic analysis to obtain an inter-annotator agreement of $0.86$ and a Cohen's Kappa score of $0.72$. In this phase we excluded $115$ papers, leaving $171$ for full-text review.

\textbf{Full-Text Review and Final Selection:}  
We assessed methodological rigor, empirical validation, and contributions to privacy-aware propaganda detection. Studies focusing on DP, federated learning, anonymization, and adversarial robustness were prioritized. Papers lacking substantial discussions on privacy, empirical analysis, or clear methodologies were excluded. Full-text screening involved three researchers, with a fourth researcher resolving disputes in ambiguous cases. In this phase we excluded nine papers, resulting in a final dataset of $162$ papers. 

\section{PROMPT Framework}

\begin{figure*}[!htp]
\centering
\includegraphics[width=0.7\linewidth]{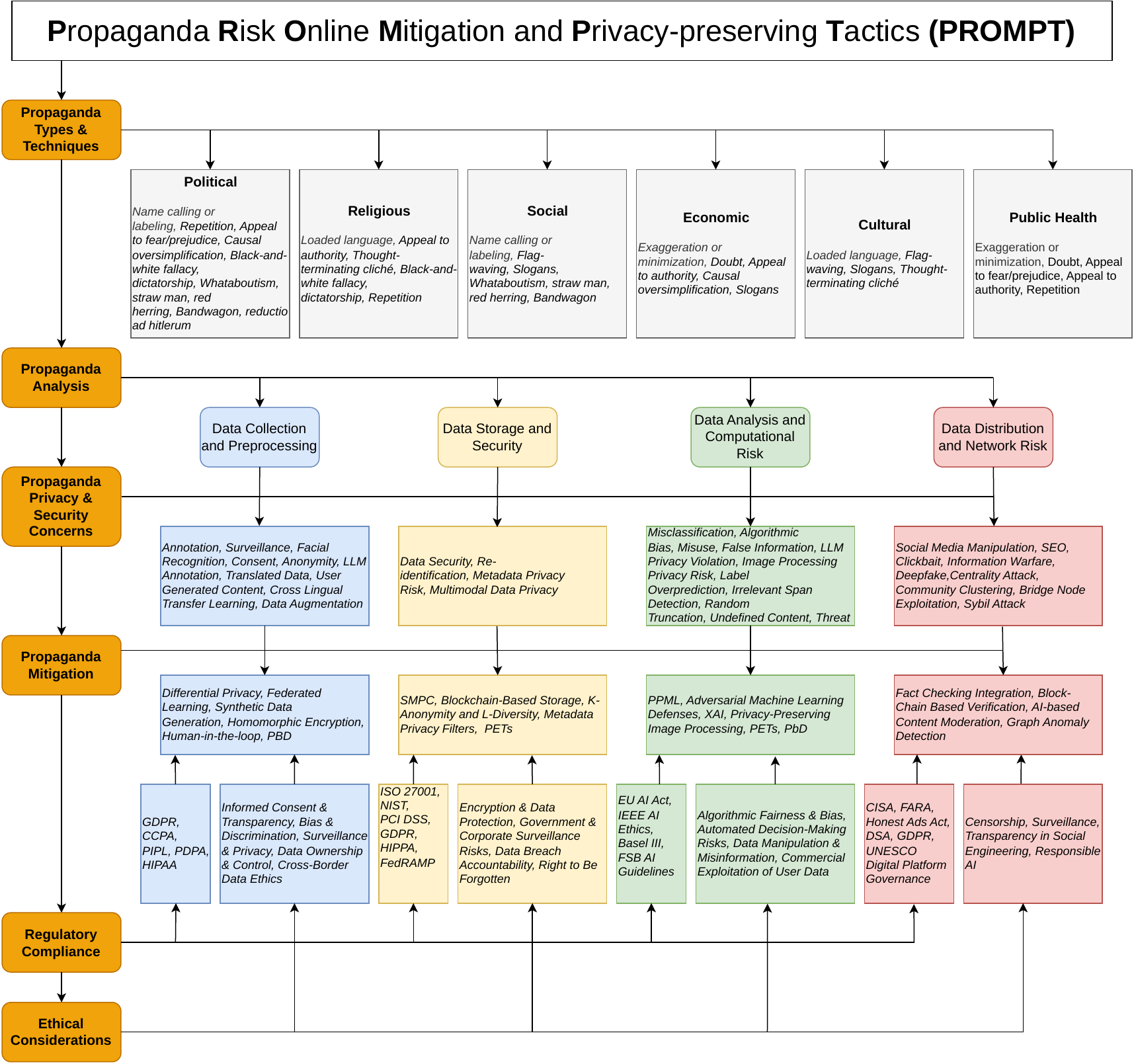}
\caption{PROMPT Framework Highlighting Privacy Risks and Mitigation Strategies}
\Description{PROMPT framework for privacy risks and Mitigation strategies.}
\label{fig:prompt}
\end{figure*}

We propose the PROMPT framework that provides a structured methodology for analyzing propaganda detection systems under privacy, security, and compliance constraints. It integrates six core components:~\textit{Propaganda Types and Techniques, Propaganda Analysis, Propaganda Privacy and Security Concerns, Propaganda Mitigation, Regulatory Compliance, and Ethical Considerations}. Unlike prior descriptive taxonomies, PROMPT formalizes the connection between propaganda techniques, computational risks, and privacy-preserving defenses. It draws on and extends prior work: the Cross-Cultural Privacy Framework~\cite{ur2013cross} that informs PROMPT's treatment of legal and cultural constraints, NLP-PRISM~\cite{goswami2026nlp} that describes privacy vulnerabilities with framework integration of different tasks related to social media, the OVERRIDE framework~\cite{raghavan2012override} that motivates its data handling mechanisms,  and PRAF~\cite{saka2024evaluating} that reinforces the integration of regulatory compliance and ethical oversight. Together, these elements define a comprehensive methodology that links detection pipelines to measurable privacy risks and mitigation strategies. Figure~\ref{fig:prompt} illustrates the PROMPT architecture, and we detail its components below.

\paragraph{\textbf{Propaganda Types and Techniques:}}
~\label{para:techniques}
We categorize propaganda into political, religious, social, economic, cultural, and public health domains each defined by recurring rhetorical and linguistic strategies~\cite{martino2020semeval}. Let $T=\{t_1, t_2, \dots, t_k\}$ denote the set of techniques (e.g., name-calling, repetition, flag-waving). For each domain $D_j$, we define a coverage score, $\mathrm{Coverage}(D_j) = \frac{|T_{D_j}|}{|T|}$,
where $T_{D_j}\subseteq T$ is the set of techniques observed in that domain and $|T|$ is the number of total number of domains defined by~\cite{da2020evaluation}. This formulation enables quantitative comparison across domains, providing measurable features that can be embedded into detection models. For example, political propaganda exhibits high technique diversity ($|T_{D_j}|$ large), while public health propaganda relies more heavily on a smaller subset such as exaggeration or fear appeals. Within PROMPT, these techniques serve as the \textit{input layer} of the framework, linking observed strategies to privacy and security risks $R$, which are later mapped to mitigation strategies $S$ via the PROMPT risk-to-defense function $\mathcal{M}:R\rightarrow S$. By grounding qualitative categories in formal representations, the framework allows reproducible benchmarking and facilitates the design of machine learning pipelines that explicitly consider adversarial robustness and privacy-aware constraints.

\paragraph{\textbf{Propaganda Analysis:}}

Propaganda analysis in PROMPT is modeled as a multi-stage pipeline where each stage exposes distinct computational and privacy risks. Let $\mathcal{P}=\{p_1,p_2,\dots,p_n\}$ denote the set of propaganda instances. The pipeline can be expressed as
\[
p_i \xrightarrow{\;\mathcal{C}\;} x_i \xrightarrow{\;\mathcal{S}\;} y_i \xrightarrow{\;\mathcal{A}\;} z_i \xrightarrow{\;\mathcal{D}\;} o_i ,
\]
where $\mathcal{C}$ is data collection and preprocessing, $\mathcal{S}$ is storage and security, $\mathcal{A}$ is analysis and computational modeling, and $\mathcal{D}$ is distribution and networking. Each stage is associated with measurable risks $R=\{r_C,r_S,r_A,r_D\}$ and potential mitigations $S=\{s_C,s_S,s_A,s_D\}$ proposed by us and defined later in the framework.

\textit{Data Collection and Preprocessing} $(\mathcal{C})$ involves acquiring multimodal propaganda artifacts (text, images, video) and transforming them into feature representations $x_i$. Risks include surveillance, lack of consent, and re-identification.  
\textit{Data Storage and Security} $(\mathcal{S})$ ensures persistence and access control for feature sets $x_i \to y_i$, where threats include metadata leakage, multimodal correlation attacks, and unauthorized retrieval.  
\textit{Data Analysis and Computational Risk} $(\mathcal{A})$ applies machine learning classifiers $f_\theta(y_i)=z_i$, with vulnerabilities including algorithmic bias, adversarial manipulation, and misclassification.  
\textit{Data Distribution and Network Risk} $(\mathcal{D})$ models how propaganda outputs $z_i$ spread across networks to yield observable influence $o_i$, with threats such as Sybil attacks, community clustering, and coordinated information warfare.

This structured representation transforms analysis from a descriptive sequence into a formal pipeline with explicit risk points, allowing integration with PROMPT's mapping $\mathcal{M}:R\to S$ to evaluate and select mitigations for each stage.

\paragraph{\textbf{Propaganda Privacy Evaluation:}}
~\label{para:evaluation}
Privacy and security challenges in propaganda detection can be represented as attack surfaces that expose user data and model integrity. We define a risk vector
$\vec{R} = \langle r_{\text{ident}}, r_{\text{meta}}, r_{\text{bias}}, r_{\text{net}} \rangle$,
where each component denotes identity leakage, metadata exposure, algorithmic bias, and network manipulation based on the four core steps $(\mathcal{C,S,A,D})$ . The overall magnitude of vulnerability can be approximated as the $L_2$ norm
\[
\|\vec{R}\|_2 = \sqrt{r_{\text{ident}}^2 + r_{\text{meta}}^2 + r_{\text{bias}}^2 + r_{\text{net}}^2}.
\]

\textit{Identity and Consent Risks} occur when annotation, LLM translation, or cross-lingual transfer enable re-identification of individuals without proper consent~\cite{solopova2024check}. These risks are measurable by record-linkage accuracy against anonymized datasets.  
\textit{Metadata Risks} emerge from insecure storage or multimodal correlation~\cite{moreno2022multidimensional}. They can be quantified using the mutual information $I(M;U)$ between metadata $M$ and user identity $U$.  
\textit{Algorithmic Risks} involve misclassification, adversarial perturbations, and censorship errors~\cite{krak2024method}. Bias can be evaluated with fairness gaps such as 
\[
\Delta_{\text{EO}} = \left| P(\hat{y}=1 \mid y=1,g=0) - P(\hat{y}=1 \mid y=1,g=1) \right|,
\]
where $g$ is a sensitive attribute and the fairness gap is the absolute value of the difference between the probability of classification with and without sensitive content exposure.
\textbf{Network Risks} stem from Sybil attacks, deepfakes, and coordinated campaigns~\cite{bessi2016social}. Their severity can be modeled by an amplification ratio
\[
A = \frac{\text{Reach}_{\text{propaganda}}}{\text{Reach}_{\text{baseline}}},
\]
which measures the inflation of dissemination relative to organic spread through network. By associating each risk with a measurable quantity such as linkage accuracy, mutual information, fairness gaps, or amplification ratios, PROMPT provides a foundation for auditing vulnerabilities with quantitative metrics instead of qualitative descriptions.

\paragraph{\textbf{Propaganda Mitigation:}}
~\label{para:mitigation}
PROMPT models mitigation as a multi-objective optimization problem that balances privacy, accuracy, and system cost. Let $R=\{r_1,\dots,r_k\}$ denote risks and $S=\{s_1,\dots,s_\ell\}$ available defenses such as DP, federated learning, SMPC, or adversarial ML defenses. For each pairing $(r_i,s_j)$ we define a mitigation utility:
\[
U(r_i,s_j) = \alpha \cdot \mathrm{PrivacyGain}(s_j) - \beta \cdot \mathrm{PerfLoss}(s_j) - \gamma \cdot \mathrm{Cost}(s_j),
\]
where $\alpha,\beta,\gamma$ are deployment-specific weights. Optimal mitigation is obtained as
\[
s^*(r_i) = \arg\max_{s_j \in S} U(r_i,s_j).
\]
This abstraction covers a wide spectrum of techniques. For example, DP increases $\mathrm{PrivacyGain}$ but induces higher $\mathrm{PerfLoss}$, while blockchain-based verification introduces computational $\mathrm{Cost}$. Integration of fact-checking, explainable AI, and human-in-the-loop review can be treated as constraints that improve contextual reliability rather than raw utility. By formalizing mitigations as optimization over $U(r_i,s_j)$, PROMPT provides a structured approach to selecting defenses that align with both security requirements and system-level trade-offs.

\paragraph{\textbf{Regulatory Compliance:}}
~\label{para:compliance}
PROMPT incorporates compliance as a measurable score that evaluates how detection pipelines adhere to legal and industry standards. Let $\mathcal{L}=\{\ell_1,\dots,\ell_m\}$ be relevant obligations (e.g., GDPR consent, CCPA opt-out, ISO 27001 security control). For a given method $f$, compliance is defined as
\[
\mathrm{CompScore}(f) = \frac{1}{m} \sum_{j=1}^{m} \mathbf{1}[f \text{ satisfies } \ell_j],
\]
where $\mathbf{1}$ is the indicator function. This allows comparison of pipelines by their degree of alignment with privacy laws, security standards, and AI governance frameworks. Beyond static measurement, compliance can be treated as a constraint in system design, requiring that selected defenses $S$ maximize utility $U(r_i,s_j)$ subject to $\mathrm{CompScore}(f) \geq \tau$, where $\tau$ is a regulatory threshold. For our analysis, we experimentally find and set to 0.319. This formulation elevates regulatory texts such as GDPR, CCPA, and the EU AI Act from descriptive guidelines into quantitative constraints that directly shape technical system choices.

\paragraph{\textbf{Ethical Considerations:}}
~\label{para:ethics}
We operationalize fairness as the proportion of safeguards relative to total ethical factors (safeguards + risks). 
Formally, let 
\[
F = \frac{|\text{safeguards}|}{|\text{safeguards}| + |\text{risks}|}, 
\quad 
B = \frac{|\text{risks}|}{|\text{safeguards}| + |\text{risks}|}.
\]
An ethically sound system minimizes $B$ while ensuring $F \geq \tau$

where $\tau$ is an empirical fairness threshold inferred from the corpus (see Table~\ref{tab:ethics-fairness}). 
In practice, this framing corresponds to enforcing informed consent during data collection, limiting surveillance, preserving user ownership of generated content~\cite{horne2023ethical}, and implementing encryption and audit protocols at the storage level~\cite{Bjola_2018}. 
At the model level, bias mitigation and explainability methods reduce discriminatory outcomes~\cite{crothers2019towards}, while at the distribution level, safeguards include transparency reporting, countermeasures against censorship, and responsible governance of automated moderation~\cite{shahriari2017ieee}.\\

% \paragraph{\textbf{Threat Model:}}
% In PROMPT we assume adversaries $\mathcal{A}$ that include state agencies, political organizations, corporations, malicious online communities, and insiders. The attack surface spans data collection, storage, analysis, and distribution. Key threats are re-identification, metadata leakage, poisoned training data, adversarial examples, deepfakes, and coordinated network attacks. Each threat $a_i \in \mathcal{A}$ is assigned likelihood $P(a_i)$ and impact $I(a_i)$. The expected system risk is
% \[
% \mathrm{Risk} = \sum_i P(a_i) \cdot I(a_i).
% \]
% Mitigation integrates technical defenses such as DP, federated learning, homomorphic encryption, privacy-preserving ML, blockchain verification, and fact checking, combined with compliance obligations under GDPR, CCPA, PIPL, and the EU AI Act. This approach reduces both technical and regulatory exposure.

\subsection{Threat Model}
In PROMPT we assume adversaries $\mathcal{A}$ characterized as follows:

\textbf{Adversary Types:} (i) Platform providers with full access to raw user data and model pipelines, (ii) external observers or third parties with black-box or API-level access, and (iii) insiders or malicious communities with partial access to data, annotations, or system outputs.
    
\textbf{Capabilities:} Adversaries may leverage data access, model querying, training data manipulation, and auxiliary information such as public profiles, external datasets, or cross-platform signals.
    
\textbf{Attack Surface:} Data collection, storage, analysis, and distribution stages of the pipeline.
    
\textbf{Privacy Violations:} Membership inference, re-identification, attribute inference and user profiling, metadata leakage, poisoned training data, adversarial examples, deepfakes, and coordinated network attacks.

Each threat $a_i \in \mathcal{A}$ is assigned likelihood $P(a_i)$ and impact $I(a_i)$. The expected system risk is $\sum_i P(a_i) \cdot I(a_i)$.

Mitigation integrates technical defenses such as DP, FL, HE, privacy-preserving ML, blockchain verification, and fact checking, combined with compliance obligations under GDPR, CCPA, PIPL, and the EU AI Act. This approach reduces both technical and regulatory exposure.

\subsection{Framework Integration}
PROMPT integrates six layers into a pipeline: \textit{types and techniques, analysis, privacy and security concerns, mitigation, compliance, and ethics}, as illustrated in Figure~\ref{fig:prompt}. Let $\mathcal{X}$ denote the propaganda inputs, $\vec{R}$ the quantified risk vector, and $S$ the set of candidate defenses. The framework defines a mapping $\mathcal{M}:\vec{R}\rightarrow S$ that selects defenses to minimize risk while satisfying compliance and fairness. Each component of $\vec{R}$ corresponds to measurable vulnerabilities - re-identification risk, metadata leakage, algorithmic bias, and network manipulation, which are explicitly linked to defenses such as DP, secure multiparty computation, or adversarially robust learning. This integration is not only conceptual (Figure~\ref{fig:prompt}) but also computational, where mitigation is expressed as an optimization problem balancing privacy gain, accuracy loss, and deployment cost. We operationalize this process in Algorithm~\ref{alg:prompt}, which details the steped procedure for estimating risks, filtering feasible defenses, scoring candidates by marginal utility, and iteratively updating the system until residual risks fall below tolerance. 

We now operationalize PROMPT's formal definitions on the collected corpus of $162$ studies. 
Specifically, the normalized risk components $r_{\text{ident}}, r_{\text{meta}}, r_{\text{bias}}, r_{\text{net}}$ and the utility function 
$U(s_j)=\alpha \cdot \mathrm{PrivacyGain}(s_j)-\beta \cdot \mathrm{PerfLoss}(s_j)-\gamma \cdot \mathrm{Cost}(s_j)$ 
are instantiated with empirical evidence from our thematic coding and compliance analysis. 
This grounding enables direct comparison between the theoretical guarantees of PROMPT and the observed fairness, bias, and performance trade-offs in transformer-based propaganda detection. 
We now turn to the Results section to quantify these outcomes.

\section{Results}

\subsection{Propaganda Types and Techniques}

\begin{table}[ht]
\centering
\caption{Evaluation of Propaganda Types and Corresponding Technique Coverage}
\resizebox{\linewidth}{!}{%
\begin{tabular}{c|c}
\hline
Propaganda Types & Propaganda Technique ($Coverage(\mathcal{D}_j)$)\\ 
\hline
Political & 0.788\\ 
Religious & 0.286\\ 
Social & 0.429\\ 
Economic & 0.286\\ 
Cultural & 0.286\\
Public health & 0.357\\ \hline
\end{tabular}%
}

\label{tab:detailed_techniques_results}
\end{table}

Propaganda detection involves analyzing types and techniques across political, religious, public health, social, economic, and cultural domains (Appendix Table~\ref{tab:results1}). 
This representation facilitates quantitative comparison of coverage and diversity across domains.

With the analysis of PROMPT framework, we find propaganda technique coverage across domains $\mathcal {D}_j$ (Table~\ref{tab:detailed_techniques_results}) using the formula from the Section ~\ref{para:techniques}. Political propaganda exhibits the highest technique coverage at 0.788, indicating greater methodological attention in detection pipelines. In contrast, religious, economic, and cultural propaganda each reflect relatively limited coverage (0.286), highlighting underexplored dimensions. Social propaganda (0.429) and public health propaganda (0.357) show moderate coverage, suggesting emerging but still incomplete methodological support.

As highlighted by Da San Martino et al.~\cite{da2019findings} and Stefan-Yurii~\cite{stefan2024cyrillic}, political propaganda employs strategies such as name-calling, repetition, black-and-white fallacy, and whataboutism to manipulate public opinion. Moreover, it includes state-sponsored efforts, electoral influence, and military-driven campaigns to sway public opinion or consolidate power~\cite{pote2024computational,kellner2019political}. Similarly, religious propaganda, as discussed by Ahmad et al.~\cite{ahmad2024semantic}, uses appeals to authority, loaded language, and thought-terminating clich{'e}s to influence belief. In the context of public health, Polonijo et al.~\cite{polonijo2021propaganda} emphasize how exaggeration, minimization, and appeals to fear shape crisis narratives. Meanwhile, social propaganda, as described by Gundapu et al.~\cite{gundapu2022automatic}, often utilizes flag-waving, slogans, and straw man arguments to polarize communities. Moreover, it involves persuasion-based techniques, false narratives, framing, bias-driven messaging, and multilingual campaigns designed to influence public sentiment and behavior~\cite{hamilton2024gpt,pandey2022detection}. Additionally, economic and cultural propaganda, as noted by Muthukumar et al.~\cite{muthukumar2024fake}, relies on causal oversimplifications, slogans, and appeals to identity-based biases. Moreover, cultural propaganda includes regional and multi-modal approaches such as Arabic memes and disinformation campaigns, which use both text and images to spread ideological messages~\cite{zaytoon2024alexunlp,abdel2023rematchka}. Finally, cross-domain techniques in propaganda detection and analysis include multilabel, multiview, and levels of analysis - such as span, sentence, and fragment-level - often employing computational models to identify and track the spread of propaganda across platforms~\cite{tomar2022araprop,chavan2022chavankane}.

Beyond conventional strategies, modern propaganda techniques have evolved to exploit digital platforms. As highlighted by Hangloo et al., social media manipulation, SEO strategies, and clickbait mechanisms are now commonly used to amplify misleading narratives. Furthermore, the increasing prevalence of deepfake content and information warfare tactics exacerbates the challenge of identifying propaganda~\cite{hangloo2022combating}. Given the growing complexity of these techniques, it is imperative to develop computational detection strategies that can adapt to domain-specific propaganda characteristics.

\subsection{Propaganda Analysis \& Evaluation}

\begin{table*}[htbp]
\centering
\caption{Evaluation of Risk Factors of Propaganda Detection.}
\renewcommand{\arraystretch}{1}
\small
\resizebox{\linewidth}{!}{%
\begin{tabular}{p{15cm} | p{0.5cm} p{0.5cm} p{0.5cm} p{0.5cm} p{0.5cm}}
\hline
\centering \textbf{Risk Factors} & $r_{\text{ident}}$
 & $r_{\text{meta}}$ & $r_{\text{bias}}$ & $r_{\text{net}}$ &
$\|\vec{R}\|_2$ \\
\hline
Cross-Lingual Transfer Learning Risk, Multimodal Data Privacy Violation, Community Clustering Exploitation, Image Processing Privacy Risk, User-Generated Content Risk, Social Media Manipulation, Irrelevant Span Detection & \riskIdent{} & - & - & - &  \riskIdent{}\\
Image Processing Privacy Risk, User-Generated Content Risk, Bridge Node Exploitation, Metadata Privacy Risk & - & \riskMeta{} & - & - & \riskMeta{} \\
LLM Privacy Violation, Data Security Breach, Label Overprediction, Information Warfare, Facial Recognition, Random Truncation, Threat Generation, Undefined Content, False Information, Centrality Attack, Misclassification, Misuse & - & - & \riskBias{} & - & \riskBias{} \\
Re-identification, Consent Violation, Algorithmic Bias, SEO Manipulation, Deepfake Attacks, Surveillance, Sybil Attack, Clickbait  & - & - & - & \riskNet{} & \riskNet{} \\

\hline
\centering \textbf{Normalized Cumulative Risk}
& & & & & \riskNorm{}\\
\hline
\end{tabular}
}

\label{tab:detailed_concerns_results}
\end{table*}

Propaganda analysis covers a comprehensive examination of how persuasive content is created, distributed, and detected in different contexts. This sub-section explores the linguistic and regional variations in this research, the digital platforms where misleading narratives proliferate, and the datasets that enable systematic study. It also reviews benchmark competitions that drive innovation and compares the computational methodologies-from traditional machine learning to cutting-edge transformer-based and multimodal approaches-used to identify and mitigate propaganda related content. By integrating these, we aim to provide a cohesive framework for understanding challenges and advancements in propaganda detection and analysis. The studies spanned across \textbf{23 languages, 18 online platforms}, mapping key sources of propaganda. Moreover, \textbf{41 datasets} were created, \textbf{5 benchmark competitions} were organized and \textbf{79 computational methodologies} were found in the literature. 

% (Appendix Figures~\ref{fig:language_distribution},~\ref{fig:platform_distribution}).

The necessity of a systematic propaganda privacy evaluation is clear, as only 5\% of surveyed studies explicitly address privacy concerns in detection pipelines ~\cite{solopova2024check,kellner2019political,solopova2023automated,syed2023hybrid,da2019findings,nabhani2025integrating}. The PROMPT framework operationalizes this evaluation across four critical stages : Data Collection and Preprocessing, Storage and Security, Analysis and Computational Risk, and Distribution and Network Risk highlighting where vulnerabilities emerge and how they accumulate. We used the formula from Section ~\ref{para:evaluation} and shown the results in Table ~\ref{tab:detailed_concerns_results}. 

\paragraph{\textbf{Data Collection and Preprocessing:}} Data collection for propaganda detection, involving textual and multimedia content from social media and news, raises concerns about surveillance, consent, and anonymity. Solopova et al. present surveillance risks from automated monitoring~\cite{solopova2024check}, while Kellner et al. highlight consent issues from collecting user data without approval~\cite{kellner2019political}. Benzmüller et al. emphasize anonymity risks in multilingual analysis~\cite{solopova2023automated}. Annotation practices may introduce systemic bias, as shown in Syed et al.’s analysis~\cite{syed2023hybrid}, while Da San Martino et al. report risks from annotation workflows themselves~\cite{da2019findings}. LLM-assisted annotation raises further concerns, with Nabhani et al. cautioning that tools like GPT-4 may inadvertently expose sensitive content~\cite{nabhani2025integrating}. 

Linguistic diversity shapes privacy risks during collection: in Europe, propaganda detection spans English~\cite{da2019findings}, French~\cite{nabhani2025integrating}, German, Dutch~\cite{golik2024dshacker}, Italian, Spanish, Polish, Greek, Russian, Georgian~\cite{lepekhin2023ftd}, Romanian~\cite{solopova2024check}, Bulgarian, North Macedonian~\cite{mahmoud2024bertastic}, Czech~\cite{herman2020propaganda}, Lithuanian~\cite{rizgeliene2024comparative}, Serbian, and Ukrainian~\cite{stefan2024cyrillic}. Conflicts such as the Russian–Ukrainian war accelerated multilingual annotation and transfer learning with human-in-the-loop checks. In North America, research focuses on English~\cite{da2019findings}, with preprocessing aimed at normalization, metadata tagging, and robustness. In South America, Spanish corpora~\cite{lepekhin2023ftd} emphasize fact-checking and sentiment alignment. Asia and the Middle East introduce further challenges, with Arabic~\cite{tomar2022araprop}, Hebrew, Hindi~\cite{nabhani2025integrating}, Urdu~\cite{ahmad2024semantic}, Chinese~\cite{zhang2022cross}, and code-mixed Hinglish/Tenglish~\cite{gundapu2022automatic}. Low-resource African languages rely on transfer learning and crowdsourcing~\cite{nabhani2025integrating}. 

Online platforms compound risks at this stage: Twitter~\cite{tomar2022araprop}, Facebook~\cite{li20211213li}, Instagram, Pinterest~\cite{zaytoon2024alexunlp}, YouTube~\cite{gundapu2022automatic}, WhatsApp~\cite{gundapu2022automatic}, Telegram~\cite{solopova2023automated}, Skype~\cite{fabrocini2021electoral}, Al Jazeera, BBC Arabic, CNN Arabic~\cite{sabol2022propaganda}, Reddit~\cite{cavaliere2023propaganda}, the dark web~\cite{last2023online}, Taobao~\cite{xing2025research}, Sina-Weibo~\cite{williamson2019trends}, and LinkedIn all introduce vulnerabilities. Datasets also contribute to privacy risks: BABE~\cite{spinde2021neural}, PTC~\cite{martino2020semeval}, CheckThat~\cite{golik2024dshacker}, PROPANEWS~\cite{huang2022faking}, AG-News~\cite{ahmad2024semantic}, TweetSpin~\cite{vijayaraghavan2022tweetspin}, DIP-ROMATS~\cite{modzelewski2024bilingual}, ArPro~\cite{hasanain2024can}, ARATWEET~\cite{mittal2022iitd},  Ar-DAD~\cite{raj2023review},  H-Prop-News~\cite{chaudhari2023empowering}, Memotion~\cite{gundapu2022automatic}, and the Meme Propaganda Techniques Corpus~\cite{chen2024multimodal}. 

Competitions such as SemEval~\cite{martino2020semeval}, WANLP~\cite{alam2022overview}, IberLEF~\cite{moral2023overview,moral2024overview}, NLP4IF~\cite{da2019findings}, and ArAIEval~\cite{hasanain-etal-2023-araieval,hasanain-etal-2024-araieval} illustrate annotation challenges but often lack transparency and consent. Among 29 privacy risks $(|\mathcal{P}|=\riskTotal)$, we found \riskIdentCount{} concerns related to this specific step $(\mathcal{C}=\riskIdentCount)$, yielding a component $r_{\text{ident}}=\riskIdent{}$.

\paragraph{\textbf{Data Storage and Security:}} Once collected, propaganda data presents long-term storage risks. Moral et al. highlight threats from storing diplomatic tweets~\cite{moral2024overview}, Modzelewski et al. show re-identification risks~\cite{modzelewski2024bilingual}, and Moreno et al. document metadata-related exposures~\cite{moreno2022multidimensional}. Mittal et al. stress that multimodal storage may disclose cultural and personal traits~\cite{mittal2022iitd}, while Semantha et al. warn of breaches from ignoring Privacy by Design~\cite{semantha2023pbdinehr}. Auñón et al. note misuse risks when datasets are redistributed without PETs~\cite{aunon2024evaluation}. Regional frameworks shape these risks: GDPR in Europe and North America enforces anonymization, Asia and the Middle East require encryption for sensitive content, while weaker regulations in South America and Africa heighten retention and repurposing risks. 

Datasets include ARATWEET~\cite{mittal2022iitd}, ArPro~\cite{hasanain2024can}, FigNews~\cite{solla2024sahara}, FaceForensics++~\cite{rossler2019faceforensics++}, and DeeperForensics~\cite{jiang2020deeperforensics}. Benchmark competitions (WANLP~\cite{alam2022overview}, IberLEF~\cite{moral2023overview}, ArAIEval~\cite{hasanain-etal-2024-araieval}) exacerbate issues by redistributing multilingual corpora without consistent anonymization. Multimodal detection magnifies vulnerabilities, as biometric and contextual signals persist long after collection. Out of the 29 identified privacy risks $(|\mathcal{P}|=29)$, only 4 $(\mathcal{S}=4)$ related to this stage have been addressed in existing studies, with the associated risk $r_{\text{meta}}$ calculated as \riskMeta{}.

\paragraph{\textbf{Data Analysis and Computational Risk:}} Analyzing propaganda introduces risks of misclassification, label overprediction, span errors, and synthetic misinformation. PROMPT emphasizes algorithmic bias, adversarial vulnerability, explainability deficits, and misinformation amplification as key threats. Early methods (BoW, TF-IDF, LDA)~\cite{da2019findings,li2019detection,moreno2022multidimensional,raj2023review} lacked nuance. Later classifiers (SVM, LR, NB, Decision Trees)~\cite{mahmoud2024bertastic,aggarwal2019nsit,patil2020bpgc,martino2020semeval,ermurachi2020uaic1860,nakov2023explainable,sabol2023augmenting} and ensembles (RF, GBDT, XGBoost)~\cite{dao2020ynu,sabol2022propaganda} improved robustness but sacrificed interpretability. Deep learning (LSTM/GRU~\cite{patil2020bpgc,arsenos2020ntuaails,kr2024method,tiessen2023privacy}, CNNs~\cite{da2019findings,martino2020semeval}, Capsule/Multi-Granularity~\cite{muthukumar2024fake,mittal2022iitd}) enhanced performance but added risk. 

Transformers (BERT, RoBERTa, mBERT, XLM-R)~\cite{da2019findings,li20211213li,gundapu10detection,grigorev2020inno}, domain-specific variants (BERTweet, DeHateBERT~\cite{casavantes2024propaltl,chavan2022large}), and generative models (BART, XLNet~\cite{huang2022faking}) advanced contextual analysis but increased bias, hallucination, and synthetic misinformation. Vision-based models (ResNet, VGG, Inception)~\cite{gundapu10detection}, multimodal architectures (CLIP, VisualBERT, UNITER~\cite{chen2024multimodal}), and large LLMs (GPT-2/3/4, LLaMA, Mixtral~\cite{morio2020hitachi,hamilton2024gpt,nabhani2025integrating,roll2024greybox}) further magnified privacy concerns due to memorization, regurgitation, and inference leakage. Tools like VADER, SHAP, Botometer~\cite{polonijo2021propaganda,pote2024computational}, and graph/ reinforcement learning methods~\cite{tundis2020mixed} extended analytical reach but remained limited by bias and efficiency trade-offs. Of the total 29 privacy risks $(|\mathcal{P}|=29)$, literature coverage exists for just 12 $(\mathcal{A}=12)$ at this step, corresponding to a risk component $r_{\text{bias}}$ of \riskBias{}.

\paragraph{\textbf{Data Distribution and Network Risk:}} At distribution, propaganda risks are amplified by network manipulation. Wijenayake et al. show clustering enables exploitation~\cite{wijenayake2025advancing}, Bessi et al. document coordinated inauthentic behavior~\cite{bessi2016social}, and Williamson et al. highlight platform vulnerabilities~\cite{williamson2019trends}. Nabhani et al. extend these risks to multilingual contexts~\cite{nabhani2025integrating}, while Wu et al. show recommendation-driven amplification~\cite{wu2022propaganda}. Modzelewski et al. caution about linguistic transfer privacy risks~\cite{modzelewski2024bilingual}, and Krak et al. point to deepfake-driven narrative shaping~\cite{krak2024method}. 

Solopova et al. warn of surveillance in distribution channels~\cite{solopova2024check}. Goodman~\cite{goodman2017honest} and Hartmann~\cite{hartmann1910mapping} highlight regulatory and historical perspectives on influence manipulation. Together, these findings show that distribution amplifies manipulation and privacy risks. Among the 29 potential privacy risks $(|\mathcal{P}|=29)$, only 10 $(\mathcal{D}=10)$ are considered in current research for this phase, resulting in a risk score $r_{\text{net}} = \riskNet{}$.

Operationalizing these risk factors  (Table~\ref{tab:detailed_concerns_results}) shows that cumulative risks across $\mathcal{C,S,A,D}$ dimensions result in $|\|\vec{R}\|_2 = 0.606$, highlighting substantial vulnerabilities. This systematic evaluation underscores the need for privacy-preserving ML, adversarial defense, and explainable AI to maintain ethical, secure, and culturally adaptive propaganda detection.

\subsection{Propaganda Mitigation} 

\begin{table}[!h]
\centering
\caption{\small Evaluation of Propaganda Mitigation ($\alpha = \beta = 0.5, \gamma = 0$)}
\resizebox{\linewidth}{!}{%
\begin{tabular}{c|cccc}
\hline
Privacy Analysis & PrivacyGain($S_j$)& PerfLoss($S_j$)& U($r_i,s_j$)& \\ 
\hline
$\mathcal{C}$ & 0.421 & 0.189 & 0.116 & \\ 
$\mathcal{S}$ & 0.526 & 0.216 & 0.155 & \\
$\mathcal{A}$ & 0.312 & 0.147 & 0.083 & \\ 
$\mathcal{D}$ & 0.263 & 0.164 & 0.049 & \\ 
\hline
\end{tabular}%
}

\end{table}

Various studies propose privacy-preserving mitigation strategies. However, our analysis reveals that these strategies appear in only 5\% of the papers on propaganda detection (Appendix Table~\ref{tab:results3}). The evaluation of mitigation across different stages ($\mathcal{C}, \mathcal{S}, \mathcal{A}, \mathcal{D}$) using the formula from Section ~\ref{para:mitigation} is shown in Table~\ref{tab:results3}, where the optimization factors $\alpha = \beta = 0.5$ provide equal weight to privacy and performance trade-off and $\gamma = 0$ was kept to show minimal impact of cost in the evaluation. From the results, $\mathcal{S}$ yields the highest PrivacyGain ($0.526$) but also a comparatively higher PerfLoss ($0.216$), while $\mathcal{C}$ strikes a more balanced trade-off between PrivacyGain ($0.421$) and PerfLoss ($0.189$). Conversely, $\mathcal{D}$ provides the lowest PrivacyGain ($0.263$) and utility ($0.049$), highlighting the uneven distribution of benefits across mitigation stages.

Semantha et al.~\cite{semantha2023pbdinehr} highlight the importance of DP, federated learning, and homomorphic encryption to protect user data. Additionally, secure multiparty computation (SMPC) and blockchain-based storage help prevent unauthorized access and ensure data integrity~\cite{aunon2024evaluation}. These methods are crucial in safeguarding data during collaborative propaganda analysis, particularly on social media~\cite{bessi2016social,vosoughi2018spread}. Moreover, adversarial machine learning (ML) defenses play a vital role in countering misinformation manipulation~\cite{pote2024computational}. By identifying and neutralizing attacks, these defenses strengthen the resilience of detection systems. Privacy-preserving image processing techniques are also essential, particularly in mitigating facial recognition risks in multimodal propaganda detection~\cite{muthukumar2024fake}. Furthermore, human-in-the-loop models enhance contextual understanding and reduce false positives~\cite{martino2020survey}, while Explainable AI (XAI) frameworks improve transparency for regulators and researchers~\cite{wu2022propaganda,shao2018anatomy}.

In addition, graph anomaly detection and blockchain verification provide robust solutions to combat misinformation spread~\cite{zannettou2019disinformation,nenadic2019unpacking}. These methods help identify coordinated misinformation campaigns and ensure data authenticity. Furthermore, privacy-enhancing technologies (PETs), such as K-anonymity, L-diversity, and metadata privacy filters, add layers of protection to sensitive user data~\cite{modzelewski2024bilingual}. These strategies work together to safeguard both privacy and security in propaganda detection systems. Lastly, the integration of these techniques is crucial for ensuring the reliability and privacy of data used in detecting propaganda. By combining various approaches, researchers can enhance the effectiveness of detection systems while protecting against malicious manipulation and preserving user privacy~\cite{hangloo2022combating,al2018prediction}.

\subsection{Regulatory Compliance} 

\begin{table*}[htbp]
\centering
\caption{Compliance evaluation with an explicit policy threshold $\hat{\tau}_c$.}
\renewcommand{\arraystretch}{1.25}
\resizebox{\textwidth}{!}{%
\begin{tabular}{c c c c c c c}
\hline
$CompScore(f)$ & $\hat{\tau}_c$ & $\Delta = CompScore(f) - \hat{\tau}_c$ & Constraint &
Safeguards Detected & Risks Detected & Recommended Fix \\
\hline
0.194 & 0.319 & -0.125 & \textcolor{red}{Violated} &
\begin{tabular}[c]{@{}c@{}}GDPR,\\ CCPA,\\ HIPAA\end{tabular} &
\begin{tabular}[c]{@{}c@{}}Privacy breaches,\\ Misinformation exposure,\\ Targeted profiling\end{tabular} &
\begin{tabular}[c]{@{}c@{}}Add consent logging, DPIAs, and transparency reporting\\ (expected $\Delta\!+\!0.306$) $\Rightarrow$ $CompScore(f)\approx 0.500 \ge \hat{\tau}_c$\end{tabular} \\
\hline
\end{tabular}
}

\label{tab:compliance-results}
\end{table*}

\begin{table*}[htbp]
\centering
\caption{Fairness bias constraint evaluation for the thematic corpus. Values use a single consistent definition of $\hat{\tau}=\frac{\text{safeguards}}{\text{safeguards}+\text{risks}}=\frac{3}{7}=0.429$.}
\renewcommand{\arraystretch}{1.25}
\resizebox{\textwidth}{!}{%
\begin{tabular}{c c c c c c c c}
\hline
$F$ (Fairness) & $B$ (Risk) & $\hat{\tau}$ & $\Delta = F - \hat{\tau}$ & Constraint &
Safeguards Detected & Risks Detected & Recommended Fix \\
\hline
0.375 & 0.500 & 0.429 & -0.054 & \textcolor{red}{Violated} &
\begin{tabular}[c]{@{}c@{}}Algorithmic Fairness \& Bias,\\ Compliance,\\ Informed Consent\end{tabular} &
\begin{tabular}[c]{@{}c@{}}Surveillance,\\ Bias/Discrimination,\\ Data Breach,\\ Misinformation/Manipulation\end{tabular} &
\begin{tabular}[c]{@{}c@{}}Transparency and reporting (+0.125 to $F$)\\ $\Rightarrow F=0.500 \geq \hat{\tau}$\end{tabular} \\
\hline
\end{tabular}
}

\label{tab:ethics-fairness}
\end{table*}

Propaganda detection systems must operate within privacy and accountability regulations, yet only 7\% of the surveyed papers explicitly address necessary compliance requirements (Appendix Table~\ref{tab:results4})~\cite{turillazzi2023digital,you2020foreign,williamson2019trends,valls2024geopolitically,al2019justdeep}. In Europe, regulations such as the GDPR and the Digital Services Act emphasize consent, data minimization, and transparency while establishing rules for harmful content and platform liability~\cite{turillazzi2023digital}. In the U.S., the Foreign Agents Registration Act (FARA) mandates disclosure of certain foreign-linked influence campaigns~\cite{you2020foreign}, demonstrating the intersection of propaganda detection with legal oversight.

Key challenges include the risk of misclassification: false positives may suppress legitimate speech and raise due-process concerns, whereas false negatives allow harmful content to persist. For instance, bot-detection errors have led to wrongful account suspensions, highlighting the importance of human review and appeal mechanisms~\cite{williamson2019trends}. Recent approaches, such as geopolitically informed models~\cite{valls2024geopolitically} and ensemble methods like JUSTDeep~\cite{al2019justdeep}, further introduce fairness and accountability considerations, underscoring the need for compliance frameworks that document data sources, model features, evaluation procedures, and error handling.

Operationalizing the compliance evaluation (Table~\ref{tab:compliance-results}) by using the formula From Section ~\ref{para:compliance}, we found a measurable gap between detected safeguards ($I=0.500$) and observed risks ($\hat{\tau}=0.319$), yielding a shortfall $\Delta=-0.125$. This indicates that existing protections are insufficient to offset risks such as privacy breaches, misinformation exposure, and targeted profiling. The minimal corrective measure involves introducing additional safeguards, with transparency and reporting identified as the most impactful. Incorporating these measures increases $CompScore(f)$ to 0.500, thereby satisfying the regulatory constraint and demonstrating how measurable improvements can bridge the compliance gap in current propaganda detection research.

\subsection{Ethical Considerations}

Ethical concerns revolve around data sensitivity, freedom of expression, and the risk of manipulation, yet only 6\% of surveyed works explicitly address such issues (Appendix Table~\ref{tab:results5})~\cite{ahuir2023elirf,cuadrado2024verbanex,valls2024geopolitically,krak2024method,barivsiccontext,chernyavskiy2024unleashing}. Prior studies highlight fairness, transparency, and user agency as essential safeguards to prevent biases, misinformation, and over-policing of viewpoints~\cite{ahuir2023elirf,krak2024method,cuadrado2024verbanex,barivsiccontext,chernyavskiy2024unleashing}, showing the importance of privacy-preserving and accountable AI-driven moderation.  

The application of the fairness–bias constraint reveals a clear gap between protective measurements and risks in current propaganda detection research. While ethical concerns such as data sensitivity, freedom of expression, and potential misuse are widely acknowledged, only a fraction of studies translate these into quantifiable safeguards. As shown in Table~\ref{tab:ethics-fairness}, merely 6\% of papers address ethical dimensions, focusing primarily on fairness and transparency in cross-lingual augmentation~\cite{ahuir2023elirf}, handling of linguistic nuances~\cite{krak2024method}, or rhetorical structures linked to offensive language~\cite{chernyavskiy2024unleashing}.

By instantiating the formula from Section~\ref{para:ethics}, we obtain $F=0.375$ for fairness coverage and $B=0.500$ for risk coverage. The inferred threshold $\hat{\tau}=0.429$ (3/7) indicates that fairness safeguards must reach at least 42.9\% of categories to offset the risks observed. Since $F < \hat{\tau}$, the ethical constraint is violated, leaving a measurable shortfall of $\Delta = -0.054$. This finding empirically validates the descriptive observation that ethical safeguards remain underdeveloped relative to the risks posed by surveillance, misinformation, and bias.

Our analysis shows that the minimal corrective step is the introduction of additional safety. Among the missing categories, \emph{transparency/reporting} is identified as the most impactful, as it counters misinformation, censorship, accountability deficits. Incorporation of this increases $F$ to 0.500, thereby satisfying the fairness constraint ($F \ge \hat{\tau}$). This confirms prior theoretical calls for transparency and fairness~\cite{ahuir2023elirf,krak2024method} and also demonstrates concretely how a single safeguard can close the measurable ethics gap in current research.

\subsection{Evaluation of Cross-Domain Propaganda Detection Dataset}

\textbf{Notation.} Throughout the experiments, $\,\perturb\,$ denotes the \emph{synthetic perturbation rate} (fraction of samples with label flips and character-level edits). Differential privacy, when discussed conceptually, is parameterized separately by $\dpnoise$ and is \emph{not} interchangeable with $\perturb$.

\begin{table}[!htp]
\centering
\caption{Evaluation of Cross-Domain Propaganda Detection Datasets with Transformer Finetuning}
\resizebox{\linewidth}{!}{%
\begin{tabular}{c|ccc}
\hline
\textbf{Base Model} & \textbf{Procedure} & \textbf{Dev F1} & \textbf{Test F1}\\
\hline
& Direct Fine-tune & 0.89 & \bf 0.89\\
\cline{2-4}
& Adversarial Defense (Back Translation) & 0.84 & 0.85\\
& NER Masking & 0.87 & 0.89\\
& Perturbation (q = 0.05) & 0.83 & 0.88\\
& Perturbation (q = 0.10) & 0.77 & 0.86\\
BERT & Perturbation (q = 0.15) & 0.71 & 0.81\\
& Perturbation (q = 0.20) & 0.67 & 0.75\\
\cline{2-4}
& BT, NER, Perturbation (q = 0.05) & 0.74 & \bf 0.83\\
& BT, NER, Perturbation (q = 0.10) & 0.69 & 0.80\\
& BT, NER, Perturbation (q = 0.15) & 0.64 & 0.75\\
& BT, NER, Perturbation (q = 0.20) & 0.62 & 0.64\\
\hline
& Direct Fine-tune &  0.90 & \bf 0.90 \\
\cline{2-4}
& Adversarial Defense (Back Translation) & 0.83 & 0.86\\
& NER Masking &  0.87 & 0.90 \\
& Perturbation (q = 0.05) &  0.84 & 0.88 \\
& Perturbation (q = 0.10) &  0.78 & 0.86 \\
GPT-2 & Perturbation (q = 0.15) &  0.73 & 0.81 \\
& Perturbation (q = 0.20) &  0.68 & 0.78 \\
\cline{2-4}
& BT, NER, Perturbation (q = 0.05) &  0.74 & \bf 0.83 \\
& BT, NER, Perturbation (q = 0.10) &  0.71 & 0.80 \\
& BT, NER, Perturbation (q = 0.15) &  0.66 & 0.76 \\
& BT, NER, Perturbation (q = 0.20) &  0.64 & 0.68 \\
\hline
\end{tabular}
}

\label{tab:finetune-results}
\end{table}

We conducted a series of fine-tuning experiments on an encoder-only model (BERT) and a decoder-only model (GPT-2) using the Cross-Domain Propaganda Detection dataset~\cite{wang2020crossdomain} (Train (10755), dev (3585), test (3585) - 60/20/20 split, binary label). This choice provides sufficient coverage of modern text-based deep learning methodologies, as encoders capture contextualized representations optimized for classification tasks~\cite{devlin2019bert}, while decoders leverage generative and sequence modeling capabilities to support classification through contextual prediction and language modeling~\cite{radford2019language}. Together, they reflect dominant paradigms in computational approaches to NLP (Table~\ref{tab:finetune-results}). 

For the encoder, direct fine-tuning of BERT yielded a baseline F1 score of $0.89$. Multilingual back translation using Xhosa, Twi, Lao, Pashto, and Yoruba reduced the score slightly to $0.85$, indicating some robustness trade-offs. Named Entity Recognition (NER) masking preserved performance, with BERT maintaining its baseline of $0.89$. Synthetic perturbation was then introduced by flipping 5\% of labels and applying character-level edits to 5\% of text (denoted $q=0.05$); higher perturbation levels ($q\in\{0.10,0.15,0.20\}$) caused progressively larger F$_1$ degradation. 

For the decoder, direct fine-tuning of GPT-2 established a baseline F1 of $0.90$. Applying multilingual back translation reduced performance slightly to $0.86$. NER masking maintained the original score of $0.90$, showing negligible impact. With synthetic perturbation at $\perturb\!=\!0.05$, GPT\mbox{-}2 reached $0.88$, with higher $\perturb$ levels again causing progressive decline. Under ensemble fine-tuning with $\perturb\!=\!0.05$, GPT\mbox{-}2 achieved an F1 of $0.83$; increasing $\perturb$ led to sharper accuracy losses.

\paragraph{\textbf{Privacy Gain vs. Utility Trade-off}}
While the results demonstrate utility degradation as $\perturb$ increases, in the context of the threat model, synthetic perturbation can be interpreted as reducing an adversary’s ability to perform concrete attacks such as membership inference, re-identification, and attribute inference by introducing uncertainty in both labels and textual features. For example, increasing $\perturb$ would be expected to lower membership inference accuracy by weakening correlations between training instances and model predictions, and to hinder re-identification by obfuscating identifiable patterns. A complete evaluation of privacy gain would therefore involve measuring adversarial success rates under these attacks (e.g., attack accuracy or AUC under standard setups such as shadow-model-based membership inference \cite{ShokriSSS17}) as a function of q, alongside utility metrics such as F1, thereby characterizing the privacy–utility trade-off.

\subsection{Quantifying Adversarial Risk}
We instantiate the threat model on representative adversarial actions extracted from the corpus. 
Each action $a_i$ is assigned a likelihood $P(a_i)$ representing how frequently it is expected to occur in practice and an impact $I(a_i)$ capturing the severity of the privacy breach if realized. 
Both values are normalized to the range $[0,1]$ to ensure comparability across heterogeneous risks. 
Table~\ref{tab:threat-model} summarizes the assigned values and the resulting weighted contributions, with metadata re-identification and model inversion emerging as the most critical threats. 
The aggregate risk score of $R=0.98$ indicates that, even under conservative estimates, adversarial actions pose a substantial cumulative risk to propaganda detection pipelines. 
This quantification provides a concrete basis for prioritizing safeguards in later stages of the PROMPT framework.

\begin{table}[htbp]
\centering
\caption{Likelihood and Impact estimates for Representative Adversarial Actions in Propaganda Detection}
\resizebox{\linewidth}{!}{%
\begin{tabular}{p{0.55\linewidth}ccc}
\toprule
\textbf{Adversarial Action $a_i$} & $P(a_i)$ & $I(a_i)$ & $P(a_i)\cdot I(a_i)$ \\
\midrule
Metadata Re-identification        & 0.6  & 0.7  & 0.42 \\
Cross-lingual Inference Leakage   & 0.4  & 0.5  & 0.20 \\
Model Inversion Attack            & 0.3  & 0.8  & 0.24 \\
Data Poisoning in Training        & 0.2  & 0.6  & 0.12 \\
\midrule
\textbf{Aggregate Risk $R$} & & & \textbf{0.98} \\
\bottomrule
\end{tabular}
}

\label{tab:threat-model}
\end{table}

\section{Discussion}
In this study, we analyze propaganda detection and mitigation with attention to privacy, security, and ethical shortcomings. Prior work highlights how political propaganda leverages rhetorical tactics such as name-calling and repetition~\cite{da2019findings}, extends to under-studied languages and platforms with resources like DIPROMATS and H-Prop-News~\cite{modzelewski2024bilingual}, and raises concerns about surveillance and re-identification~\cite{solopova2024check}. Privacy-preserving methods including differential privacy and federated learning have been advocated~\cite{semantha2023pbdinehr}, alongside compliance with GDPR and ISO 27001~\cite{voigt2017eu} and safeguards for fairness and transparency~\cite{ahuir2023elirf}. By situating our analysis within the PROMPT framework, we address these challenges and demonstrate how privacy and ethics can be embedded in propaganda detection, guiding the evaluation framework used to answer our research questions.

\textit{\textbf{Answer to ~\hyperref[rq:1]{RQ1}:}} Propaganda detection in online platforms poses significant privacy risks due to its reliance on large-scale user-generated content. These systems collect, store, process vast amounts of data, including text, images, and user interactions, which often contain PII, metadata, and behavioral patterns~\cite{nakov2023explainable}. As a result, users become vulnerable to tracking, profiling, and re-identification, even when anonymization techniques are applied. Advanced inference methods can still expose anonymized data, leading to unauthorized surveillance and data exploitation~\cite{solopova2023automated}. Furthermore, machine learning models trained on user data are susceptible to model inversion attacks, where adversaries can reconstruct original inputs, potentially revealing sensitive information~\cite{modzelewski2024bilingual}. 

Algorithmic profiling presents additional concerns, as it may categorize users based on ideological stances, leading to discrimination, censorship, or unintended biases in content moderation~\cite{krak2024method}. Inadequate encryption and weak access controls further exacerbate privacy threats, increasing the likelihood of data breaches. Moreover, the diversity of global legal frameworks complicates regulatory compliance, creating inconsistencies in data protection enforcement across jurisdictions~\cite{fabrocini2021electoral}. Beyond technical risks, ethical concerns arise, particularly in politically sensitive contexts. Automated propaganda detection mechanisms could be misused to suppress dissenting voices or disproportionately target marginalized groups, undermining freedom of expression and democratic discourse~\cite{hasanain-etal-2023-araieval}. We have found \textbf{0.606} normalized cumulative privacy risk factor across all the steps.

\textit{\textbf{Answer to ~\hyperref[rq:2]{RQ2}:}} Various techniques assess and mitigate privacy risks in propaganda detection; however, each has significant limitations. In particular, privacy risk assessment often involves differential privacy, federated learning, and adversarial testing to evaluate data exposure and model vulnerabilities. For instance, DP measures leakage risk by adding noise to training data, which helps reduce exposure. Nevertheless, this approach degrades model accuracy, making it a less optimal solution in certain contexts~\cite{gaanoun2022si2m}. Similarly, federated learning minimizes centralized data collection by decentralizing training, thereby reducing direct access to sensitive data. Yet, it remains vulnerable to gradient inversion attacks, which can reconstruct sensitive information from model updates~\cite{moral2024overview}.In addition to these methods, traditional anonymization techniques, such as entity removal and content masking, assess privacy by identifying sensitive text. However, these techniques often fail to prevent re-identification, especially when attackers utilize auxiliary data. Furthermore, they tend to reduce detection accuracy, thereby limiting their overall effectiveness~\cite{oliinyk2020propaganda}. Likewise, adversarial testing is used to evaluate how models handle manipulated inputs; yet, it struggles to keep up with evolving attack strategies, making it an imperfect solution~\cite{sabol2022propaganda}. Beyond text-based methods, network-based assessments are employed to detect manipulation patterns, such as Sybil and centrality attacks. While these techniques enhance detection capabilities, metadata exposure remains a significant concern~\cite{al2018prediction,krak2024method}. 

Additionally, multimodal detection, which integrates text, audio, and visual data, improves accuracy; however, it also introduces privacy risks through metadata cross-referencing, potentially exposing user-sensitive information~\cite{cui2023multimodal,wu2022propaganda}. Finally, large language models (LLMs) pose further challenges in assessing privacy risks due to their black-box nature, making their security vulnerabilities harder to evaluate. Although these techniques help mitigate privacy risks to some extent, they require trade-offs between privacy protection and propaganda detection effectiveness. We observe a privacy–utility trade-off (0.049–0.155) under synthetic perturbation ($\perturb$), with up to a 7\% performance drop in privacy-preserved transformer finetuning.

\textit{\textbf{Answer to ~\hyperref[rq:3]{RQ3}:}} Privacy-preserving strategies play a crucial role in enhancing the security and integrity of propaganda detection models. One effective approach is applying differential privacy at the feature extraction stage, which perturbs high-risk features while maintaining detection performance~\cite{zaytoon2024alexunlp}. Additionally, secure federated learning, when combined with SMPC and homomorphic encryption, ensures that model updates remain protected from inference attacks~\cite{laskar2022cnlp}. Moreover, privacy-preserving techniques such as tokenization-based de-identification, adversarial text sanitization help mask sensitive details while preserving linguistic coherence, making them valuable for mitigating privacy risks~\cite{nabhani2025integrating}. To further enhance transparency and accountability, XAI methods, including SHAP and LIME, facilitate risk assessment and bias detection, thus improving trust in the detection models~\cite{fouad2024sussexai}. At the same time, strengthening adversarial robustness through gradient masking and perturbation defenses fortifies models against manipulation attempts, ensuring their reliability in adversarial settings~\cite{da2019findings}. 

Furthermore, automated privacy audits play a key role in complying with regulations such as GDPR~\cite{goldman2020introduction} and CCPA~\cite{laskar2022cnlp}, helping organizations align with legal requirements. In politically sensitive contexts, integrating Human-in-the-Loop (HITL) mechanisms offers an additional layer of oversight, reducing risks associated with automated decision-making~\cite{zaytoon2024alexunlp}. Finally, the PROMPT framework serves as a robust solution for online propaganda analysis. Through its six-step propaganda analysis, PROMPT provides a comprehensive approach, reinforcing privacy-preserving measures and improving security and ethical integrity. By PROMPT analysis, we have found a lack of \textbf{0.306} in compliance score and \textbf{0.125} in fairness threshold which states that existing regulatory compliance and ethical aspect should be more agile for privacy and security preservation in this domain.

% \begin{resultbox}[Key Findings:]
% \small Privacy risks and protective measurements in propaganda detection are unevenly distributed across the pipeline: most work concentrates on model-level defenses while data collection, storage, and distribution stages remain largely unprotected. Multilingual and multimodal settings show the highest vulnerability, with limited adoption of privacy-enhancing technologies and weak regulatory alignment. These imbalances suggest that progress in propaganda detection depends not only on better models but also on end-to-end integration of privacy and compliance controls across all stages of the pipeline.
% \end{resultbox}

\section{Implications}
Grounding our synthesis of 162 studies in the PROMPT framework, we identify system-level implications for deploying privacy-aware and regulation-aligned propaganda detection. Current pipelines exhibit a normalized privacy risk of $\lVert \vec{R} \rVert_2 = 0.606$ (Tables~\ref{tab:compliance-results},~\ref{tab:ethics-fairness}), limited adoption of privacy-enhancing technologies, and measurable compliance and fairness deficits. These observations demonstrate that existing systems inherit structural weaknesses across all pipeline stages rather than isolated failures at individual modules. As a result, end-to-end redesign is necessary to align detection practices with emerging regulatory, ethical, and technical expectations.

\subsection[Data Collection and Preprocessing (C)]{Data Collection and Preprocessing $(\mathcal{C})$}
Data acquisition continues to drive the highest concentration of privacy exposure, particularly in multilingual and multimodal environments where inference pathways expand rapidly. Re-identification, covert profiling, and consent violations occur when pipelines aggregate sensitive linguistic cues or cross-reference behavioral metadata. Automated entity redaction, provenance tracking, data minimization, and dataset nutrition labels reduce downstream exposure and establish transparent, auditable data lineage~\cite{cummings2020individual}. Federated curation and differential privacy adapters replace centralized retention in high-sensitivity contexts, thereby reducing attack surface and limiting the blast radius of potential breaches~\cite{el2022differential}. Large language model-based annotation workflows require formal leakage analysis because prompts, intermediate reasoning traces, and labels frequently encode identifiable information~\cite{he2024emerged}. Treating annotation as a controlled computational activity rather than an informal step promotes reproducibility, meets regulatory expectations for traceability, and reduces the likelihood of unintentional data leaks.

\subsection[Storage and Security (S)]{Storage and Security $(\mathcal{S})$}
Metadata correlation, multimodal linkability, and unrestricted redistribution increase long-term privacy risk even when primary content appears sanitized. Encryption at rest, granular access control, and retention limits form a defensible baseline, while metadata minimization and controlled hashing restrict inference channels that enable re-identification~\cite{waqar2013framework}. Machine-readable usage policies, audit logs, and Data Protection Impact Assessments elevate pipeline transparency and equip external auditors with verifiable evidence of compliance. Integrating compliance scoring into development and release workflows, with automatic blocking when scores fall below the threshold $\hat{\tau}_c$, enforces regulatory alignment and prevents the deployment of systems that violate GDPR, CCPA, or the EU AI Act. These mechanisms transform compliance from a post hoc audit activity into a continuous, measurable design constraint.

\subsection[Analysis and Modeling (A)]{Analysis and Modeling $(\mathcal{A})$}
Model development remains disproportionately centered on accuracy, leaving privacy, robustness, and fairness underexamined. This imbalance produces pipelines that perform well under benchmark conditions but introduce substantial risk during real-world deployment. Reporting protocols that pair accuracy with robustness under perturbation, cross-group fairness gaps, and explicit privacy–utility trade-offs provide a multidimensional view of system behavior and reveal risks that single-metric evaluation conceals~\cite{shokri2015privacy}. Privacy-preserving training techniques such as DP-SGD, federated optimization, encrypted inference, and privacy-preserving multimodal joins enhance protection in high-risk domains~\cite{qu2020privacy}. Explainability techniques supply structured evidence for debugging, model validation, and regulatory assessment, enabling investigators to trace harmful outputs and identify structural biases. Mapping each component of $\vec{R}$ to a mitigation strategy in $M: R \rightarrow S$ embeds principled defense selection into the development cycle, replacing ad hoc adjustments with systematic risk reasoning.

\subsection[Distribution and Networks (D)]{Distribution and Networks $(\mathcal{D})$}
Propagation networks amplify both privacy risk and societal harm, especially when platforms rely on graph-based features to detect coordinated behavior. These features expose relational patterns that enable adversarial inference and user correlation. Noise injection into graph embeddings, subgraph sampling, and private anomaly detection maintain sensitivity to coordinated manipulation while reducing the exposure of relational structure~\cite{roman2013features}. Transparency reporting, user-notification channels, and structured appeal pathways strengthen procedural accountability and reduce false-positive harms during moderation. Ensuring that deployment occurs only when fairness and compliance conditions satisfy $F \geq \hat{\tau}$ prevents the release of models that disproportionately affect vulnerable communities or fail to satisfy jurisdictional regulatory thresholds. These controls operationalize fairness and compliance as measurable gates rather than aspirational ideals.

\subsection{Operational Usage of PROMPT}
The PROMPT framework functions as an operational playbook that unifies risk quantification, mitigation selection, and compliance evaluation into a single iterative cycle. It establishes a structured procedure in which practitioners identify dominant risks, map them to feasible defenses, and refine the pipeline until residual risk falls within organizational tolerance. This cycle parallels mature security engineering workflows by linking qualitative threat modeling with quantitative measures that guide decision-making at each stage of system development. Operational use of PROMPT begins with estimating the risk vector $\vec{R}$ and identifying components that exceed acceptable thresholds. The mapping $M: R \rightarrow S$ then selects mitigation strategies that reduce risk while respecting performance and cost constraints. As updates are incorporated, practitioners track $(\lVert \vec{R} \rVert_2, \mathrm{CompScore}, F)$ across successive releases to evaluate how technical changes influence privacy exposure, regulatory alignment, and fairness guarantees. Documenting marginal utility effects $U(r_i, s_j)$ provides an auditable record that links each intervention to measurable reductions in risk, creating transparent justification for design choices~\cite{shahriar2023survey}.

This structure elevates privacy, fairness, and compliance from aspirational goals to operational metrics embedded directly into development and governance pipelines. Rather than applying controls after deployment, teams integrate them throughout the lifecycle, from dataset design to distribution. The result is a reproducible process that exposes trade-offs early, supports independent verification, and strengthens organizational accountability. By enforcing continuous monitoring and transparent reporting, the PROMPT playbook enables propaganda detection systems to function as measurable, trustworthy, and regulation-aligned socio-technical infrastructures that maintain resilience under evolving adversarial and policy pressures~\cite{alshammari2018model}.

\subsection{State Power and Information Asymmetry}
Beyond system-level considerations, propaganda detection operates within broader geopolitical and institutional power asymmetries. State actors and large platforms possess disproportionate access to data, computational resources, and regulatory influence, enabling large-scale monitoring, narrative shaping, and cross-border information control. While privacy-preserving mechanisms in PROMPT reduce risks such as re-identification, membership inference, and user profiling, they may also limit transparency and accountability when applied in high-stakes moderation or governance contexts. This creates an inherent trade-off between protecting individual privacy and enabling oversight of powerful actors engaged in coordinated information operations. In this setting, the deployment of privacy-aware propaganda detection must be carefully balanced to avoid reinforcing existing asymmetries. Mechanisms such as auditable pipelines, transparency reporting, and independent evaluation become critical to ensure that privacy protections do not obscure harmful large-scale manipulation. Situating PROMPT within these dynamics highlights the need for governance frameworks that jointly address privacy, accountability, and the geopolitical implications of controlling information flows.

\section{Limitations and Future Work}
In this study, we surveyed a broad spectrum of privacy-preserving techniques and highlighted their role in propaganda detection, but we did not evaluate their effectiveness in operational deployments. We focused our corpus on top-tier conferences and journals, which allowed us to capture state-of-the-art research while inevitably leaving out some relevant work from other venues and industry reports. In future work, we will expand our dataset to cover non-English and under-represented sources, with particular attention to low-resource contexts. We also plan to conduct empirical experiments that measure the integration costs and performance trade-offs of privacy-enhancing technologies when deployed at scale.

\section{Conclusion}
Online propaganda poses risks that demand detection pipelines which are accurate, privacy-preserving, and regulation-compliant, yet advances in ML and DL, while improving detection accuracy, introduce new vulnerabilities such as metadata exposure, adversarial inference, compliance gaps. In this work, we present the first systematic evaluation of privacy, fairness, and compliance in propaganda detection, analyzing $162$ peer-reviewed publications through the PROMPT framework, which we developed. Our analysis showed that 81\% of systems rely on sensitive user data while only 18\% adopt privacy-preserving technologies, yielding a normalized corpus risk of $\|\vec{R}\|_2=0.606$. We further demonstrated that fairness coverage consistently falls short of the required threshold and that compliance deficits of $0.306$ leave current pipelines ethically and legally fragile. Empirical experiments established concrete baselines for the privacy–utility trade-off, with privacy constraints causing up to 22\% performance degradation. Finally, our thematic mapping exposed uneven attention to propaganda techniques and a lack of controls and measurements in multilingual and multimodal settings.By operationalizing risk vectors, compliance scores, and fairness thresholds, we turned ethical and legal considerations into quantifiable requirements, establishing a rigorous basis for evaluating and improving propaganda detection systems.

%-------------------------------------------------------------------------------

% optional clearing of the page

\section*{Ethical Considerations}

This SoK 
% examines privacy and security risks in propaganda detection systems by mapping vulnerabilities, synthesizing mitigation strategies, and guiding future research toward privacy-preserving, ethical solutions. We 
use only publicly available datasets and examples, without collecting new social media data, interacting with users, or processing personally identifiable information. Given the sensitivity of propaganda tied to political, cultural, religious, and demographic identities, we avoid deanonymization, real-world adversarial attacks, and any design of manipulation or disinformation systems; all attack vectors (e.g., re-identification, metadata leakage, adversarial misinformation) are discussed conceptually. 
% The PROMPT framework centers ethical and regulatory concerns for vulnerable communities and points out gaps in privacy, security, and fairness to promote transparent, accountable, and harm-preventive system design.
\begin{acks}
We would like to acknowledge the Data Agency and Security (DAS) Lab at George Mason University. This work is partially funded by Google Research Awards. The opinions expressed in this work are solely those of the authors. 
\end{acks}
% % optional clearing of the page


\bibliographystyle{ACM-Reference-Format}
\bibliography{sample-base}
%\vfill\break
\appendix

\section*{Artefacts}

\noindent PROMPT framework integration and defense selection is shown Algorithm~\ref{alg:prompt}.

\begin{algorithm}[!htp]
\small
\caption{PROMPT Integration and Defense Selection}
\small
\label{alg:prompt}
\begin{algorithmic}[1]
\Require Inputs $\mathcal{X}$, techniques $T$, stages $(\mathcal{C},\mathcal{S},\mathcal{A},\mathcal{D})$, mitigations $S$, obligations $\mathcal{L}$, thresholds $\tau_c,\tau_e$, budget $B$, tolerance $\epsilon$, min improv.\ $\eta$, max iters $K$
\Ensure Defenses $\mathcal{S}^\star$, updated model $f'$, audit $\mathcal{A}_{\text{udit}}$

\State \textbf{Featurize} from $T$: compute $\mathrm{Coverage}(D_j)$ and embed
\State \textbf{Init risk}
\Statex \quad $r_{\text{ident}}\!\gets\!\mathrm{LinkageAcc}$,\ $r_{\text{meta}}\!\gets\!I(M;U)$,\ $r_{\text{bias}}\!\gets\!\Delta_{\mathrm{EO}}$,\ $r_{\text{net}}\!\gets\!A$
\Statex \quad $\vec{R}\gets\langle r_{\text{ident}},r_{\text{meta}},r_{\text{bias}},r_{\text{net}}\rangle$, $C\gets0$, $f'\gets f$, $\mathcal{S}^\star\gets\varnothing$, $k\gets0$

\While{$\|\vec{R}\|_2>\epsilon$ \textbf{and} $k<K$}
  \State $k\gets k+1$
  \State \textbf{Filter feasible} candidates
  \Statex \quad $\mathcal{C}_{\text{and}}\gets \{s\!\in\!S:\ C+\mathrm{Cost}(s)\le B,\ \mathrm{CompScore}(f')\!\ge\!\tau_c,\ F(f')\!\ge\!\tau_e\}$
  \If{$\mathcal{C}_{\text{and}}=\varnothing$} \textbf{break} \Comment no feasible action \EndIf

  \State \textbf{Score each} $s\in\mathcal{C}_{\text{and}}$
  \ForAll{$s\in\mathcal{C}_{\text{and}}$}
     \State $\Delta R(s)\gets \|\vec{R}\|_2-\|\vec{R}-\mathrm{Mitigate}(s)\|$  \Comment risk reduction
     \State $U(s)\gets \alpha\,\Delta R(s)-\beta\,\mathrm{PerfLoss}(s)-\gamma\,\mathrm{Cost}(s)$
  \EndFor

  \State $s^\star \gets \arg\max_{s\in\mathcal{C}_{\text{and}}}\big(U(s),\ \Delta R(s)/\max\{1,\mathrm{Cost}(s)\}\big)$
  \If{$U(s^\star)\le0$ \textbf{or} $\Delta R(s^\star)<\eta$} \textbf{break} \Comment no beneficial action \EndIf

  \State \textbf{Apply} $s^\star$, update
  \Statex \quad $\vec{R}\gets \vec{R}-\mathrm{Mitigate}(s^\star)$,\ $C\gets C+\mathrm{Cost}(s^\star)$,\ $\mathcal{S}^\star\gets \mathcal{S}^\star\cup\{s^\star\}$
  \State \textbf{Update compliance and fairness}
  \Statex \quad $\displaystyle \mathrm{CompScore}(f')\gets \frac{1}{|\mathcal{L}|}\sum_{\ell\in\mathcal{L}}\mathbf{1}[f'\ \text{satisfies}\ \ell]$
  \Statex \quad $F(f')\gets \text{fairness metric value}$
  \State $S\gets S\setminus\{s^\star\}$
\EndWhile

\State $\mathcal{A}_{\text{udit}}\gets\{\vec{R},\ \mathcal{S}^\star,\ C,\ \mathrm{CompScore}(f'),\ F(f'),\ \text{violated constraints}\}$
\State \Return $\mathcal{S}^\star,\ f',\ \mathcal{A}_{\text{udit}}$
\end{algorithmic}
\end{algorithm}

Detailed propaganda detection survey results of Language, Platform, Datasets, Benchmark Competition, Computational Methodology are shown in Table \ref{tab:result}.

\begin{table*}[tbp]

\centering
\caption{Survey Result of the Propaganda : Language, Platform, Datasets, Benchmark Competitions, Computational Methodology Reported for all $162$ Papers.}
\resizebox{\linewidth}{!}{%
\begin{tabular}{|p{2cm}|p{23.5cm}|}
\hline
\textbf{Key Points} & \textbf{Survey Results}\\
\hline
\textbf{Language} &  \textbf{Asia:} Chinese~\cite{zhang2022cross,waight2021chinese,xing2025research,ahmad2024semantic},
Hindi~\cite{nabhani2025integrating,fouad2024sussexai,lepekhin2023ftd,alhabashi2024asos,shah2024mememind,solla2024sahara,ragab2025multilingual,alhindi2019fine,gundapu2022automatic,aggarwal2019nsit,arsenos2020ntuaails,verma2020transformers,kim2020ttui,jiang2020umsiforeseer,chaudhari2023empowering},
Urdu~\cite{ahmad2024semantic,ahmad2023-Pro-propaganda},
Code-Mixed (Hinglish, Tenglish-Telugu-English, Hindi-English mix)~\cite{gundapu2022automatic}

\textbf{Middle East and Africa:}
Arabic~\cite{tomar2022araprop,nabhani2025integrating,refaee2022arabem,chavan2022chavankane,laskar2022cnlp,hussein2022ngu,attieh2022pythoneers,fouad2024sussexai,mohtaj2022tub,zaytoon2024alexunlp,abdel2023rematchka,alhabashi2024asos,mahmoud2024bertastic,roll2024greybox,shah2024mememind,solla2024sahara,ragab2025multilingual,aldabbas2025multiprop,sharara2022arabert,hasanain2024can,mittal2022iitd,chavan2022large,xiao2023nexus,alam2022overview,golik2024dshacker,aggarwal2019nsit,arsenos2020ntuaails,gaanoun2022si2m,verma2020transformers,kim2020ttui,jiang2020umsiforeseer,last2023online,ahmad2024semantic,richardsuse},
Hebrew~\cite{nabhani2025integrating,fouad2024sussexai,lepekhin2023ftd,alhabashi2024asos,shah2024mememind,solla2024sahara,ragab2025multilingual,aldabbas2025multiprop,solopova2024check,kellner2019political,kellner2020s,chernyavskiy2024unleashing,pezo2024all,aggarwal2019nsit,arsenos2020ntuaails,verma2020transformers,kim2020ttui,jiang2020umsiforeseer,purificato2023apatt}

\textbf{Europe and America:}
English~\cite{hou2019caunlp,li2019detection,da2019findings,morio2020hitachi,fadel2019pretrained,li20211213li,hou2021fpai,kaczynski2021homados,nabhani2025integrating,fouad2024sussexai,lepekhin2023ftd,alhabashi2024asos,mahmoud2024bertastic,roll2024greybox,shah2024mememind,solla2024sahara,ragab2025multilingual,aldabbas2025multiprop,aggarwal2019nsit,mapes2019divisive,vlad2019sentence,tian2021mind,patil2020bpgc,solopova2024check,pote2024computational,blaschke2020cyberwalle,gundapu10detection,alhindi2019fine,grigorev2020inno,sprenkamp2023large,wang2020leveraging,gupta2019neural,martino2020semeval,raj2020solomon,kranzlein2020team,saleh2019team,mikhalkova2020utmn,dao2020ynu,moreno2022multidimensional,gundapu2022automatic,modzelewski2024bilingual,golik2024dshacker,tian2023efficient,ahuir2023elirf,rodrigo2023hierarchical,oliinyk2020propaganda,casavantes2024propaltl,casavantes2023propaltl,fernandez2024victor,al2019justdeep,valls2024geopolitically,hamilton2024gpt,tundis2020mixed,zavolokina2024think,chernyavskiy2024unleashing,barfar2022linguistic,tundis2023detection,syed2023hybrid,pezo2024all,puvselj2020propaganda,achimescu2020feeding,solopova2023automated,nakov2023explainable,vorakitphan2022protect,aggarwal2019nsit,arsenos2020ntuaails,chauhan2020psuedoprop,dementieva2020skoltechnlp,verma2020transformers,kim2020ttui,ermurachi2020uaic1860,jiang2020umsiforeseer,tiwari2023lstm,dewantara2020combination,zhang2022cross,pandey2022detection,maseri2020socio,cui2023multimodal,polonijo2021propaganda,wu2022propaganda,purificato2023apatt,moral2023overview,moral2024overview,lytvyn2024enhancing,szwoch2024limitations,vijayaraghavan2022tweetspin,ahmad2025hierarchical,ahmad2023-Pro-propaganda,yu2021interpretable,franklin2020russian,cuadrado2024verbanex,przybyla2023does,rajmohan2022emotion,chen2024multimodal,last2023online,cavaliere2023propaganda,ahmad2024semantic,hamilton2021towards,richardsuse,barivsiccontext},
Bulgarian~\cite{mahmoud2024bertastic,roll2024greybox,stefan2024cyrillic},
Czech~\cite{herman2020propaganda,sabol2022propaganda,sabol2023augmenting,baisa2019benchmark},
Dutch~\cite{golik2024dshacker},
French~\cite{nabhani2025integrating,fouad2024sussexai,lepekhin2023ftd,alhabashi2024asos,shah2024mememind,solla2024sahara,ragab2025multilingual,aldabbas2025multiprop,solopova2024check,chernyavskiy2024unleashing,pezo2024all,solopova2023automated,aggarwal2019nsit,arsenos2020ntuaails,verma2020transformers,kim2020ttui,jiang2020umsiforeseer,purificato2023apatt},
German~\cite{lepekhin2023ftd,aldabbas2025multiprop,solopova2024check,kellner2019political,kellner2020s,chernyavskiy2024unleashing,pezo2024all,purificato2023apatt},
Georgian~\cite{lepekhin2023ftd,pezo2024all},
Greek~\cite{pezo2024all,lepekhin2023ftd},
Italian~\cite{lepekhin2023ftd,pezo2024all,aldabbas2025multiprop,solopova2024check,chernyavskiy2024unleashing,vorakitphan2022protect,purificato2023apatt,fabrocini2021electoral},
Lithuanian~\cite{rizgeliene2024comparative},
North Macedonian~\cite{mahmoud2024bertastic,roll2024greybox},
Polish~\cite{lepekhin2023ftd,pezo2024all,aldabbas2025multiprop,chernyavskiy2024unleashing,purificato2023apatt,szwoch2024limitations},
Romanian~\cite{solopova2024check,solopova2023automated},
Russian~\cite{lepekhin2023ftd,aldabbas2025multiprop,aggarwal2019nsit,solopova2024check,kellner2019political,kellner2020s,chernyavskiy2024unleashing,stefan2024cyrillic,pezo2024all,solopova2023automated,purificato2023apatt,hein2023propaganda,franklin2020russian,last2023online},
Serbian~\cite{stefan2024cyrillic},
Spanish~\cite{lepekhin2023ftd,modzelewski2024bilingual,ahuir2023elirf,rodrigo2023hierarchical,casavantes2024propaltl,casavantes2023propaltl,fernandez2024victor,valls2024geopolitically,pezo2024all,moral2023overview,moral2024overview,cuadrado2024verbanex,last2023online},
Ukrainian~\cite{solopova2024check,kr2024method,krak2025method,stefan2024cyrillic,solopova2023automated,krak2024method}\\
\hline
\textbf{Platform} & \textbf{Social Media:} 
Twitter~\cite{tomar2022araprop,refaee2022arabem,chavan2022chavankane,laskar2022cnlp,attieh2022pythoneers,mohtaj2022tub,abdel2023rematchka,aldabbas2025multiprop,sharara2022arabert,pote2024computational,chavan2022large,xiao2023nexus,alam2022overview,kellner2019political,moreno2022multidimensional,ahuir2023elirf,rodrigo2023hierarchical,casavantes2023propaltl,fernandez2024victor,al2019justdeep,valls2024geopolitically,kellner2020s,gaanoun2022si2m,raj2023review,zhang2022cross,maseri2020socio,moral2023overview,moral2024overview,szwoch2024limitations,vijayaraghavan2022tweetspin,ahmad2025hierarchical,ahmad2023-Pro-propaganda,williamson2019trends,rajmohan2022emotion,last2023online,cavaliere2023propaganda,casavantes2023propitter,ellermann2016terror},  
Facebook~\cite{li20211213li,hou2021fpai,kaczynski2021homados,nabhani2025integrating,zaytoon2024alexunlp,alhabashi2024asos,mahmoud2024bertastic,shah2024mememind,ragab2025multilingual,tian2021mind,gundapu2022automatic,maseri2020socio,cui2023multimodal,wu2022propaganda,ahmad2025hierarchical,franklin2020russian,chen2024multimodal,last2023online,cavaliere2023propaganda,fabrocini2021electoral},  
Instagram~\cite{zaytoon2024alexunlp,alhabashi2024asos,gundapu2022automatic},  
Pinterest~\cite{zaytoon2024alexunlp,alhabashi2024asos},  
YouTube~\cite{gundapu2022automatic,ellermann2016terror}  

\textbf{News Media:}
Al Jazeera~\cite{alam2022overview},  
BBC Arabic~\cite{alam2022overview},  
CNN Arabic~\cite{alam2022overview},  
Sky News Arabia~\cite{alam2022overview},  
Czech News Websites~\cite{sabol2022propaganda,sabol2023augmenting}  

\textbf{Discussion Forums:} 
Reddit~\cite{cavaliere2023propaganda},  
Dark Web Forum~\cite{last2023online}  

\textbf{Messaging:}  
WhatsApp~\cite{gundapu2022automatic,fabrocini2021electoral},  
Telegram~\cite{solopova2023automated},  
Skype~\cite{fabrocini2021electoral}  

\textbf{Other Digital Platforms:} 
Taobao~\cite{xing2025research},  
Sina-Weibo~\cite{williamson2019trends}\\
\hline

\textbf{Datasets} & \textbf{English:}  
AG-News~\cite{ahmad2024semantic}, BABE~\cite{rodrigo2023hierarchical}, COVID-19 Fake News Dataset~\cite{gundapu2022automatic}, CheckThat~\cite{golik2024dshacker}, MH17 Tweets Dataset~\cite{al2019justdeep}, DARPA Twitter Bot Challenge~\cite{williamson2019trends}, IRA Dataset~\cite{last2023online}, Kaggle Fake News Dataset~\cite{last2023online}, MBIC~\cite{rodrigo2023hierarchical}, PTC (Propaganda Techniques Corpus)~\cite{li2019detection,da2019findings,morio2020hitachi,fadel2019pretrained,aggarwal2019nsit,mapes2019divisive,vlad2019sentence,martino2020survey,patil2020bpgc,blaschke2020cyberwalle,wang2020leveraging,martino2020semeval,raj2020solomon,hamilton2024gpt,szwoch2024limitations,ahmad2025hierarchical,przybyla2023does,ahmad2024semantic,hamilton2021towards}, PROPANEWS Dataset~\cite{huang2022faking}, PONC~\cite{szwoch2024limitations}, ProSoul~\cite{ahmad2023-Pro-propaganda}, ProText~\cite{ahmad2025hierarchical,ahmad2023-Pro-propaganda,ahmad2024semantic}, QProp~\cite{martino2020survey,ahmad2025hierarchical,ahmad2024semantic}, THSP 17~\cite{martino2020survey}, TweetSpin~\cite{vijayaraghavan2022tweetspin}  

\textbf{Arabic:}  
ARATWEET~\cite{mittal2022iitd}, Ar-DAD~\cite{raj2023review}, ArPro~\cite{hasanain2024can}, IED~\cite{ahmad2023-Pro-propaganda}

\textbf{Multilingual:}  
DIPROMATS~\cite{modzelewski2024bilingual,tian2023efficient,ahuir2023elirf,rodrigo2023hierarchical,casavantes2024propaltl,casavantes2023propaltl,fernandez2024victor,al2019justdeep,valls2024geopolitically,moral2023overview,moral2024overview,cuadrado2024verbanex}, MultiProp~\cite{aldabbas2025multiprop}, H-Prop-News Dataset~\cite{chaudhari2023empowering}, Czech Propaganda Benchmark Dataset~\cite{sabol2022propaganda}, Chinese and English Propaganda Detection Dataset~\cite{zhang2022cross}, Sogou~\cite{ahmad2024semantic}, Propitter~\cite{casavantes2023propitter}, SentiMix~\cite{gundapu2022automatic} 

\textbf{Multimodal:}  
Celeb-DF~\cite{raj2023review}, DeeperForensics, Deepfake Forensics~\cite{raj2023review}, Deepfake TIMIT~\cite{raj2023review}, FaceForensics~\cite{raj2023review}, FaceForensics++~\cite{raj2023review}, UADFV~\cite{raj2023review}, Vid-TIMIT~\cite{raj2023review}, Memotion Analysis Dataset~\cite{gundapu2022automatic}, Meme Propaganda Techniques Corpus~\cite{chen2024multimodal}, FigNews~\cite{solla2024sahara,ragab2025multilingual}\\
\hline
\textbf{Benchmark Competition} & SemEval~\cite{morio2020hitachi,li20211213li,hou2021fpai,kaczynski2021homados,lepekhin2023ftd,mahmoud2024bertastic,roll2024greybox,tian2021mind,patil2020bpgc,blaschke2020cyberwalle,grigorev2020inno,martino2020semeval,raj2020solomon,kranzlein2020team,saleh2019team,mikhalkova2020utmn,dao2020ynu,aggarwal2019nsit,arsenos2020ntuaails,chauhan2020psuedoprop,dementieva2020skoltechnlp,verma2020transformers,kim2020ttui,ermurachi2020uaic1860,jiang2020umsiforeseer,purificato2023apatt}, WANLP~\cite{tomar2022araprop,refaee2022arabem,chavan2022chavankane,laskar2022cnlp,hussein2022ngu,attieh2022pythoneers,mohtaj2022tub,sharara2022arabert,mittal2022iitd,chavan2022large,alam2022overview,gaanoun2022si2m}, IberLEF~\cite{modzelewski2024bilingual,tian2023efficient,ahuir2023elirf,rodrigo2023hierarchical,casavantes2024propaltl,casavantes2023propaltl}, NLP4IF~\cite{hou2019caunlp,da2019findings,aggarwal2019nsit,alhindi2019fine,al2019justdeep,hamilton2024gpt,vorakitphan2022protect,ahmad2023-Pro-propaganda}, ArAIEval~\cite{nabhani2025integrating,fouad2024sussexai,zaytoon2024alexunlp,abdel2023rematchka,alhabashi2024asos,shah2024mememind,xiao2023nexus}\\
\hline
\textbf{Computational Methodology} & \textbf{Statistical ML:} BoW~\cite{da2019findings}, Decision Trees~\cite{nakov2023explainable,sabol2023augmenting,williamson2019trends,last2023online}, Gradient-Boosted Decision Trees (GBDT)~\cite{sabol2023augmenting}, KNN~\cite{moral2023overview,last2023online}, LDA~\cite{moreno2022multidimensional,raj2023review,maseri2020socio,last2023online}, LIWC~\cite{alhindi2019fine}, Logistic Regression~\cite{patil2020bpgc,martino2020semeval,raj2020solomon,saleh2019team,mikhalkova2020utmn,moreno2022multidimensional,oliinyk2020propaganda,tundis2020mixed,achimescu2020feeding,last2023online}, Naive Bayes~\cite{aggarwal2019nsit,ermurachi2020uaic1860,sabol2022propaganda,baisa2019benchmark,last2023online}, PCA~\cite{raj2023review}, Random Forest~\cite{martino2020semeval,sabol2022propaganda,franklin2020russian,williamson2019trends}, Recursive Binary Split (RBS) classification tree~\cite{franklin2020russian}, SVM~\cite{mahmoud2024bertastic,aggarwal2019nsit,solopova2024check,raj2020solomon,tundis2020mixed,solopova2023automated,ermurachi2020uaic1860,raj2023review,zhang2022cross,maseri2020socio,vijayaraghavan2022tweetspin,baisa2019benchmark,last2023online}, TF-IDF~\cite{li2019detection,aggarwal2019nsit,ahmad2023propaganda,oliinyk2020propaganda,cuadrado2024verbanex}, XGBoost~\cite{patil2020bpgc,dao2020ynu,raj2023review,zhang2022cross}, Linear Regression~\cite{solopova2023automated}

\textbf{Deep Learning:} Bi-LSTM~\cite{vlad2019sentence,blaschke2020cyberwalle,dao2020ynu,aggarwal2019nsit,arsenos2020ntuaails,dementieva2020skoltechnlp,chaudhari2023empowering}, CNN~\cite{da2019findings,patil2020bpgc,gupta2019neural,martino2020semeval,moreno2022multidimensional,tundis2020mixed,dementieva2020skoltechnlp,verma2020transformers,raj2023review,dewantara2020combination,maseri2020socio,chaudhari2023empowering,ahmad2025hierarchical,rajmohan2022emotion,chen2024multimodal},CapsuleNet~\cite{muthukumar2024fake}, ConvoNet~\cite{polonijo2021propaganda}, DenseNet121~\cite{muthukumar2024fake}, ELMo~\cite{aggarwal2019nsit,aggarwal2019nsit,arsenos2020ntuaails,verma2020transformers,ahmad2024semantic}, GANs~\cite{raj2023review}, GRU~\cite{kr2024method,aggarwal2019nsit,tiessen2023privacy},LSTMs~\cite{patil2020bpgc,arsenos2020ntuaails}, Multi-Granularity Network~\cite{mittal2022iitd}, RNN~\cite{moreno2022multidimensional,raj2023review,pandey2022detection,maseri2020socio,krak2024method,ahmad2023-Pro-propaganda},SCST~\cite{huang2022faking},SpanNER~\cite{przybyla2023does}

\textbf{Transformers:} ALBERT~\cite{gundapu10detection,grigorev2020inno,wu2022propaganda,purificato2023apatt,ahmad2025hierarchical}, AraBERT~\cite{nabhani2025integrating,laskar2022cnlp,attieh2022pythoneers,fouad2024sussexai,mohtaj2022tub,abdel2023rematchka,sharara2022arabert,chavan2022large,alam2022overview,gaanoun2022si2m}, AraELECTRA~\cite{chavan2022large}, AraGPT~\cite{abdel2023rematchka}, BART~\cite{huang2022faking}, BERT~\cite{li2019detection,da2019findings,fadel2019pretrained,kaczynski2021homados,refaee2022arabem,aggarwal2019nsit,mapes2019divisive,vlad2019sentence,gundapu10detection,alhindi2019fine,grigorev2020inno,gupta2019neural,alam2022overview,martino2020semeval,raj2020solomon,kranzlein2020team,moreno2022multidimensional,stefan2024cyrillic,achimescu2020feeding,solopova2023automated,vorakitphan2022protect,dementieva2020skoltechnlp,verma2020transformers,jiang2020umsiforeseer,cui2023multimodal,wu2022propaganda,purificato2023apatt,chaudhari2023empowering,ahmad2025hierarchical,yu2021interpretable,hein2023propaganda,przybyla2023does,xing2025research,chen2024multimodal,barivsiccontext}, BERTweet~\cite{casavantes2024propaltl,casavantes2023propaltl}, DeHateBERT~\cite{chavan2022large}, DistilBERT~\cite{grigorev2020inno,purificato2023apatt,cavaliere2023propaganda}, HerBERT~\cite{purificato2023apatt}, MARBERT~\cite{chavan2022large,alam2022overview,golik2024dshacker}, mBERT~\cite{ragab2025multilingual,mittal2022iitd,zhang2022cross}, mT5~\cite{ragab2025multilingual}, RoBERTa~\cite{li20211213li,ragab2025multilingual,aggarwal2019nsit,patil2020bpgc,gundapu10detection,grigorev2020inno,chavan2022large,alam2022overview,martino2020semeval,raj2020solomon,golik2024dshacker,rodrigo2023hierarchical,vorakitphan2022protect,zhang2022cross,purificato2023apatt,sabol2022propaganda,moral2023overview,ahmad2025hierarchical,ahmad2023-Pro-propaganda,hein2023propaganda,cuadrado2024verbanex,kelkar2024multimodal,ahmad2024semantic}, RoBERTuito~\cite{casavantes2024propaltl,casavantes2023propaltl}, Robeczech~\cite{sabol2022propaganda}, TwHIN-BERT~\cite{valls2024geopolitically}, XLM-R~\cite{hussein2022ngu,ragab2025multilingual,mittal2022iitd,chavan2022large,alam2022overview,martino2020semeval,golik2024dshacker,rodrigo2023hierarchical,zhang2022cross,cuadrado2024verbanex}, XLNet~\cite{patil2020bpgc,gundapu10detection,grigorev2020inno,martino2020semeval,purificato2023apatt,ahmad2025hierarchical}

\textbf{Vision:} CLIP~\cite{chen2024multimodal}, DALLE2~\cite{shah2024mememind}, InceptionV3~\cite{gundapu10detection}, NF-ResNet50~\cite{wu2022propaganda}, ResNet-152~\cite{gundapu10detection,chen2024multimodal}, ResNet101~\cite{cui2023multimodal}, ResNet50~\cite{li20211213li,alhabashi2024asos,wu2022propaganda}, UNITER~\cite{gundapu10detection}, VGG-19~\cite{gundapu10detection}, VLM~\cite{mahmoud2024bertastic}, ViLBERT~\cite{gundapu10detection}, VisualBERT~\cite{gundapu10detection,wu2022propaganda}, YOLO-CNN~\cite{raj2023review}

\textbf{LLMs:} GPT-2~\cite{morio2020hitachi,aldabbas2025multiprop,patil2020bpgc}, GPT-3~\cite{hamilton2024gpt,stefan2024cyrillic,szwoch2024limitations}, GPT-3.5~\cite{hamilton2024gpt,stefan2024cyrillic,szwoch2024limitations}, GPT-4~\cite{nabhani2025integrating,hasanain2024can,hamilton2024gpt,zavolokina2024think,stefan2024cyrillic,achimescu2020feeding,szwoch2024limitations}, GROVER~\cite{huang2022faking}, Mistral~\cite{lytvyn2024enhancing}, Mixtral~\cite{roll2024greybox},LLaMa2~\cite{roll2024greybox}, Chain of Thought~\cite{sprenkamp2023large}, LM-BFF~\cite{wu2022propaganda}

\textbf{Other Techniques:} Botometer~\cite{pote2024computational}, HDSF~\cite{huang2022faking}, LatexPRO~\cite{wang2020leveraging}, MViTO-GAT~\cite{chen2024multimodal}, OCR~\cite{tian2021mind,tundis2020mixed}, Reinforcement Learning~\cite{martino2020survey}, VADER~\cite{polonijo2021propaganda}\\
\hline

\end{tabular}%
}

\label{tab:result}

\end{table*}

Statistical results are shown in Tables ~\ref{tab:results1}, ~\ref{tab:results2}, ~\ref{tab:results3}, ~\ref{tab:results4}, ~\ref{tab:results5}.

\begin{table}[H]

\centering
\caption{Propaganda Types and Techniques.}
\renewcommand{\arraystretch}{1}

\resizebox{\linewidth}{!}{%
\begin{tabular}{p{3cm} | p{0.15cm} p{0.15cm} p{0.15cm} p{0.15cm} p{0.15cm} p{0.15cm} p{0.15cm} p{0.15cm} p{0.15cm} p{0.15cm} p{0.15cm} p{0.15cm} p{0.15cm} p{0.15cm}}
\hline
\multicolumn{1}{c}{} & \multicolumn{14}{c} {\textbf{Propaganda Techniques}} \\
\hline
\textbf{Propaganda Types} &  
 \rotatebox{90}{Whataboutism, Straw man, Red herring} & 
 \rotatebox{90}{Bandwagon, Reductio ad Hitlerum} & 
 \rotatebox{90}{Exaggeration, minimization} & 
 \rotatebox{90}{Thought-terminating clich{\'e}} & 
 \rotatebox{90}{Causal oversimplification} & 
 \rotatebox{90}{Appeal to fear/prejudice} & 
 \rotatebox{90}{Black-and-white fallacy} & 
 \rotatebox{90}{Name calling, labeling} & 
 \rotatebox{90}{Appeal to authority} & 
 \rotatebox{90}{Loaded Language} & 
 \rotatebox{90}{Flag-waving} & 
 \rotatebox{90}{Repetition} & 
 \rotatebox{90}{Slogans} & 
 \rotatebox{90}{Doubt} \\
\hline
Political~\cite{da2019findings,da2020evaluation,stefan2024cyrillic} 
 & $\bullet$ & $\bullet$ & $\bullet$ &  & $\bullet$ & $\bullet$ & $\bullet$ & $\bullet$ &  & $\bullet$ & $\bullet$ & $\bullet$ & $\bullet$ &  \\
 \hline
Religious~\cite{ahmad2025hierarchical} 
 &  &  &  & $\bullet$ &  &  & $\bullet$ &  & $\bullet$ & $\bullet$ &  &  &  &  \\
 \hline
Social~\cite{hamilton2024gpt,pandey2022detection,gundapu2022automatic} 
 & $\bullet$ & $\bullet$ &  &  &  & $\bullet$ &  & $\bullet$ &  &  & $\bullet$ &  & $\bullet$ &  \\
 \hline
Economic~\cite{muthukumar2024fake} 
 &  &  & $\bullet$  &  & $\bullet$  &  &  &  & $\bullet$  &  &  &  &  & $\bullet$  \\
 \hline
Cultural~\cite{zaytoon2024alexunlp,abdel2023rematchka,muthukumar2024fake} 
 &  &  &  & $\bullet$ &  &  &  &  & $\bullet$ & $\bullet$ & $\bullet$ &  &  &  \\
 \hline
Public Health~\cite{polonijo2021propaganda} 
 &  &  & $\bullet$ &  &  & $\bullet$ &  &  & $\bullet$ &  &  & $\bullet$ &  & $\bullet$ \\
 \hline
\end{tabular}
}

\label{tab:results1}
\end{table}

\begin{table}[H]

\centering
\caption{Security Concerns of Propaganda Detection.}
\renewcommand{\arraystretch}{2}
%\small
\resizebox{\linewidth}{!}{%
\begin{tabular}{p{1.75cm} | p{0.1cm} p{0.1cm} p{0.1cm} p{0.1cm} p{0.1cm} p{0.1cm} p{0.1cm} p{0.1cm} p{0.1cm} p{0.1cm} p{0.1cm} p{0.1cm} p{0.1cm} p{0.1cm} p{0.1cm} p{0.1cm} p{0.1cm} p{0.1cm} p{0.1cm} p{0.1cm} p{0.1cm} p{0.1cm} p{0.1cm} p{0.1cm} p{0.1cm} p{0.1cm} p{0.15cm} p{0.1cm} p{0.1cm}}
\hline
\multicolumn{1}{c}{} & \multicolumn{29}{c}{\textbf{Security Concerns}} \\  
\hline
\textbf{Privacy Analysis} & 
 \rotatebox{90}{Cross-Lingual Transfer Learning Risk} & 
 \rotatebox{90}{Multimodal Data Privacy Violation} & 
 \rotatebox{90}{Community Clustering Exploitation} & 
 \rotatebox{90}{Image Processing Privacy Risk} & 
 \rotatebox{90}{User-Generated Content Risk} & 
 \rotatebox{90}{Social Media Manipulation} & 
 \rotatebox{90}{Irrelevant Span Detection} & 
 \rotatebox{90}{Bridge Node Exploitation} & 
 \rotatebox{90}{Metadata Privacy Risk} & 
 \rotatebox{90}{LLM Privacy Violation} & 
 \rotatebox{90}{Data Security Breach} & 
 \rotatebox{90}{Label Overprediction} & 
 \rotatebox{90}{Information Warfare} & 
 \rotatebox{90}{Facial Recognition} & 
 \rotatebox{90}{Random Truncation} & 
 \rotatebox{90}{Threat Generation} & 
 \rotatebox{90}{Undefined Content} & 
 \rotatebox{90}{False Information} & 
 \rotatebox{90}{Centrality Attack} & 
 \rotatebox{90}{Misclassification} & 
 \rotatebox{90}{Re-identification} & 
 \rotatebox{90}{Consent Violation} & 
 \rotatebox{90}{Algorithmic Bias} & 
 \rotatebox{90}{SEO Manipulation} & 
 \rotatebox{90}{Deepfake Attacks} & 
 \rotatebox{90}{Surveillance} & 
 \rotatebox{90}{Sybil Attack} & 
 \rotatebox{90}{Clickbait} & 
 \rotatebox{90}{Misuse} \\
\hline
$\mathcal{C}$~\cite{solopova2024check,ragab2025multilingual}
 & $\bullet$ & $\bullet$ & $\bullet$ & $\bullet$ & $\bullet$ & $\bullet$ & $\bullet$
 &  &  &  &  &  &  &  &  &  &  &  &  &  &  &  &  &  &  &  &  &  \\
 \hline
$\mathcal{S}$~\cite{moreno2022multidimensional,kellner2019political}
 &  &  &  & $\bullet$ & $\bullet$
 &  &  & $\bullet$ & $\bullet$
 &  &  &  &  &  &  &  &  &  &  &  &  &  &  &  &  &  &  &  \\
 \hline
$\mathcal{A}$~\cite{krak2024method,tian2023efficient}
 &  &  &  &  &  &  &  &  &  &
   $\bullet$ & $\bullet$ & $\bullet$ & $\bullet$ & $\bullet$
 & $\bullet$ & $\bullet$ & $\bullet$ & $\bullet$ & $\bullet$ & $\bullet$
 &  &  &  &  &  &  &  & & $\bullet$ \\
 \hline
$\mathcal{D}$~\cite{williamson2019trends,nabhani2025integrating}
 &  &  &  &  &  &  &  &  &  &  &  &  &  &  &  &  &  &
   & $\bullet$  & $\bullet$  & $\bullet$  & $\bullet$  & $\bullet$ 
 & $\bullet$  & $\bullet$  & $\bullet$  & $\bullet$  & $\bullet$  \\
\hline
\end{tabular}
}

\label{tab:results2}
\end{table}

\begin{table}[H]

\centering
\caption{Propaganda Mitigation Techniques.}
\renewcommand{\arraystretch}{1.25}
\resizebox{\linewidth}{!}{%
\begin{tabular}{p{2.75cm}| p{0.15cm} p{0.15cm} p{0.15cm} p{0.15cm} p{0.15cm} p{0.15cm} p{0.15cm} p{0.15cm} p{0.15cm} p{0.15cm} p{0.15cm} p{0.15cm} p{0.15cm} p{0.15cm} p{0.15cm} p{0.15cm} p{0.15cm} p{0.15cm} p{0.15cm}}
\hline
\multicolumn{1}{c}{} & \multicolumn{19}{c}{\textbf{Mitigation Techniques}} \\  
\hline
\textbf{Privacy Analysis} & 
 \rotatebox{90}{Privacy-Preserving Machine Learning (PPML)} & 
 \rotatebox{90}{Adversarial Machine Learning Defenses} & 
 \rotatebox{90}{Secure Multi-Party Computation (SMPC)} & 
 \rotatebox{90}{Privacy-Enhancing Technologies (PETs)} & 
 \rotatebox{90}{Privacy-Preserving Image Processing} & 
 \rotatebox{90}{Blockchain-Based Verification} & 
 \rotatebox{90}{AI-Based Content Moderation} & 
 \rotatebox{90}{K-Anonymity and L-Diversity} & 
 \rotatebox{90}{Synthetic Data Generation} & 
 \rotatebox{90}{Fact-Checking Integration} & 
 \rotatebox{90}{Metadata Privacy Filters} & 
 \rotatebox{90}{Blockchain-Based Storage} & 
 \rotatebox{90}{Graph Anomaly Detection} & 
 \rotatebox{90}{Privacy by Design (PbD)} & 
 \rotatebox{90}{HE} & 
 \rotatebox{90}{XAI} & 
 \rotatebox{90}{DP} & 
 \rotatebox{90}{FL} & 
 \rotatebox{90}{HITL} \\
\hline
$\mathcal{C}$~\cite{modzelewski2024bilingual,wu2022propaganda,pote2024computational} & 
$\bullet$ &  &  & $\bullet$ &  & $\bullet$ &  &  & $\bullet$ & $\bullet$ &  &  &  & $\bullet$ &  &  & $\bullet$ &  & $\bullet$ \\
\hline
$\mathcal{S}$~\cite{semantha2023pbdinehr,aunon2024evaluation} & 
$\bullet$ & $\bullet$ & $\bullet$ &  &  & $\bullet$ & $\bullet$ & $\bullet$ &  &  & $\bullet$ & $\bullet$ &  & $\bullet$ & $\bullet$ &  &  &  &  \\
\hline
$\mathcal{A}$~\cite{muthukumar2024fake,martino2020survey} &  
 & $\bullet$ &  &  & $\bullet$ &  &  & $\bullet$ &  &  &  &  & $\bullet$ &  & $\bullet$ & $\bullet$ &  &  &  \\
 \hline
$\mathcal{D}$~\cite{wu2022propaganda,modzelewski2024bilingual} &  
 &  &  & $\bullet$ &  &  &  &  & $\bullet$ &  &  & $\bullet$ &  &  &  &  &  & $\bullet$ & $\bullet$ \\
\hline
\end{tabular}
}

\label{tab:results3}
\end{table}

\begin{table}[H]

\centering
\caption{Regulatory Compliance of Propaganda Detection.}
\renewcommand{\arraystretch}{1.25}

\resizebox{\linewidth}{!}{%
\begin{tabular}{p{2.75cm} | p{0.15cm} p{0.15cm} p{0.15cm} p{0.15cm} p{0.15cm} p{0.15cm} p{0.15cm} p{0.15cm} p{0.15cm} p{0.15cm} p{0.15cm} p{0.15cm} p{0.15cm} p{0.15cm} p{0.15cm} p{0.15cm} p{0.15cm} p{0.15cm}}
\hline
\multicolumn{1}{c}{} & \multicolumn{18}{c}{\textbf{Regulatory Compliances}} \\  
\hline
\textbf{Privacy Analysis} & 
 \rotatebox{90}{UNESCO-Digital Platform Governance} & 
 \rotatebox{90}{FSB AI Guidelines} & 
 \rotatebox{90}{IEEE AI Ethics} & 
 \rotatebox{90}{Honest Ads Act} & 
 \rotatebox{90}{Basel III} & 
 \rotatebox{90}{EU AI Act} & 
 \rotatebox{90}{ISO 27001} & 
 \rotatebox{90}{PCI DSS} & 
 \rotatebox{90}{FedRAMP} & 
 \rotatebox{90}{HIPAA} & 
 \rotatebox{90}{NIST} & 
 \rotatebox{90}{CCPA} & 
 \rotatebox{90}{PDPA} & 
 \rotatebox{90}{CISA} & 
 \rotatebox{90}{FARA} & 
 \rotatebox{90}{PIPL} & 
 \rotatebox{90}{GDPR} & 
 \rotatebox{90}{DSA} \\
\hline
$\mathcal{C}$~\cite{ahuir2023elirf,cuadrado2024verbanex,barivsiccontext,jelassi2023towards} & 
$\bullet$ &  &  &  &  &  &  &  &  &  &  & $\bullet$ &  &  &  & $\bullet$ & $\bullet$ &  \\
\hline
$\mathcal{S}$~\cite{krak2024method,williamson2019trends,you2020foreign,turillazzi2023digital} &  
 &  &  &  &  &  & $\bullet$ & $\bullet$ &  & $\bullet$ & $\bullet$ &  & $\bullet$ &  & $\bullet$ & $\bullet$ & $\bullet$ & $\bullet$ \\
 \hline
$\mathcal{A}$~\cite{valls2024geopolitically,solopova2024check,hartmann1910mapping} &  
 & $\bullet$ & $\bullet$ &  & $\bullet$ & $\bullet$ &  &  &  &  &  &  &  &  &  &  &  &  \\
 \hline
$\mathcal{D}$~\cite{krak2024method,solopova2024check,goodman2017honest} &  
$\bullet$ &  &  & $\bullet$ &  &  &  &  & $\bullet$ &  &  &  &  & $\bullet$ & $\bullet$ &  &  & $\bullet$ \\
\hline
\end{tabular}
}

\label{tab:results4}
\end{table}

\begin{table}[H]

\centering
\caption{Ethical Considerations of Propaganda Detection.}
\renewcommand{\arraystretch}{1.25}

\resizebox{\linewidth}{!}{%
\begin{tabular}{p{2.5cm} | p{0.15cm} p{0.15cm} p{0.15cm} p{0.15cm} p{0.15cm} p{0.15cm} p{0.15cm} p{0.15cm} p{0.15cm} p{0.15cm} p{0.15cm} p{0.15cm} p{0.15cm} p{0.15cm} p{0.15cm} p{0.15cm}}
\hline
\multicolumn{1}{c}{} & \multicolumn{16}{c}{\textbf{Ethical Considerations}} \\  
\hline
\textbf{Privacy Analysis} & 
 \rotatebox{90}{Government \& Corporate Surveillance Risks} & 
 \rotatebox{90}{Commercial Exploitation of User Data} & 
 \rotatebox{90}{Data Manipulation \& Misinformation} & 
 \rotatebox{90}{Transparency in Social Engineering} & 
 \rotatebox{90}{Informed Consent \& Transparency} & 
 \rotatebox{90}{Automated Decision-Making Risks} & 
 \rotatebox{90}{Encryption \& Data Protection} & 
 \rotatebox{90}{Algorithmic Fairness \& Bias} & 
 \rotatebox{90}{Data Breach Accountability} & 
 \rotatebox{90}{Data Ownership \& Control} & 
 \rotatebox{90}{Cross-Border Data Ethics} & 
 \rotatebox{90}{Surveillance \& Privacy} & 
 \rotatebox{90}{Bias \& Discrimination} & 
 \rotatebox{90}{Right to Be Forgotten} & 
 \rotatebox{90}{Responsible AI} & 
 \rotatebox{90}{Censorship} \\
\hline
$\mathcal{C}$~\cite{ahuir2023elirf,cuadrado2024verbanex,valls2024geopolitically} &  
 &  & $\bullet$ & $\bullet$ & $\bullet$ &  &  &  &  & $\bullet$ & $\bullet$ & $\bullet$ &  &  &  &  \\
 \hline
$\mathcal{S}$~\cite{krak2024method,williamson2019trends} &  
$\bullet$ &  & $\bullet$ &  &  &  & $\bullet$ &  & $\bullet$ & $\bullet$ &  & $\bullet$ &  &  &  &  \\
\hline
$\mathcal{A}$~\cite{barivsiccontext,hartmann1910mapping} &  
 & $\bullet$ & $\bullet$ &  &  & $\bullet$ &  & $\bullet$ &  &  &  &  & $\bullet$ &  & $\bullet$ & $\bullet$ \\
 \hline
$\mathcal{D}$~\cite{solopova2024check,hartmann1910mapping} &  
 & $\bullet$ & $\bullet$ & $\bullet$ &  &  &  & $\bullet$ &  &  & $\bullet$ &  & $\bullet$ & $\bullet$ & $\bullet$ & $\bullet$ \\
\hline
\end{tabular}
}

\label{tab:results5}
\end{table}

\end{document}